\RequirePackage{fix-cm}
\documentclass{article}                     
\usepackage{graphicx}
\graphicspath{ {./images/} }
\usepackage{xcolor}
\definecolor{sapphirecrayola}{rgb}{0.18,0.36,0.63}
\definecolor{aoenglish}{rgb}{0.0,0.5,0.0}
\usepackage{amssymb}
\usepackage{amsmath}
\usepackage{nccmath}
\usepackage{mathtools}
\DeclarePairedDelimiter\ceil{\lceil}{\rceil}
\usepackage{subfig}
\usepackage{placeins}
\usepackage{afterpage}
\edef\svtheparindent{\the\parindent}
\usepackage{parskip}
\parindent=\svtheparindent\relax
\usepackage{setspace}
\usepackage{enumitem}
\usepackage{comment}
\usepackage{hyperref}
\usepackage{pdfpages}

\usepackage{natbib}
\setcitestyle{authoryear,open={(},close={)}}
\newcommand\eqdef{\mathrel{\overset{\makebox[0pt]{\mbox{\normalfont\tiny\sffamily def}}}{=}}}
\begin{document}
\title{Data Analytics Driven Controlling: bridging statistical modeling and managerial intuition \thanks{This research was supported by the Deutsche Forschungsgesellschaft through the International Research Training Group 1792 "High Dimensional Nonstationary Time Series". This research is based upon work from COST Action 19130, supported by COST (European Cooperation in Science and Technology). Danial Saef acknowledges the support of PwC.}
}


\author{Kainat Khowaja\thanks{Humboldt-Universität zu Berlin, International Research Training Group 1792, Spandauer Str. 1, 10178 Berlin, Germany. Email: kainat.khowaja@hu-berlin.de} \and 
        Danial Saef\thanks{Humboldt-Universität zu Berlin, International Research Training Group 1792, Spandauer Str. 1, 10178 Berlin, Germany. PwC, Germany. Email: danial.saef@pwc.com} \and
        Sergej Sizov\thanks{PricewaterhouseCoopers, Germany. Email: sergej.sizov@pwc.com}   \and    Wolfgang Karl Härdle\thanks{Humboldt-Universität zu Berlin, BRC Blockchain Research Center, Berlin; Sim Kee Boon Institute, Singapore Management University, Singapore; WISE Wang Yanan Institute for Studies in Economics, Xiamen University, Xiamen, China; National Chiao Tung University, Dept Information Science and Finance, Hsinchu, Taiwan, ROC; Charles University, Dept Mathematics and Physics, Prague, Czech Republic; Grant CAS: XDA 23020303 and DFG IRTG 1792 gratefully acknowledged. Email: haerdle@hu-berlin.de} 
}
\date{}
\maketitle
\begin{abstract}Strategic planning in a corporate environment is often based on experience and intuition, although internal data is usually available and can be a valuable source of information. Predicting merger \& acquisition (M\&A) events is at the heart of strategic management, yet not sufficiently motivated by data analytics driven controlling. One of the main obstacles in using e.g. count data time series for M\&A seems to be the fact that the intensity of M\&A is time varying at least in certain business sectors, e.g. communications. We propose a new automatic procedure to bridge this obstacle using novel statistical methods. The proposed approach allows for a selection of adaptive windows in count data sets by detecting significant changes in the intensity of events. We test the efficacy of the proposed method on a simulated count data set and put it into action on various M\&A data sets. It is robust to aberrant behaviour and generates accurate forecasts for the evaluated business sectors. It also provides guidance for an a-priori selection of fixed windows for forecasting. Furthermore, it can be generalized to other business lines, e.g. for managing supply chains, sales forecasts, or call center arrivals, thus giving managers new ways for incorporating statistical modeling in strategic planning decisions.

\end{abstract}
\clearpage
\doublespacing

\section{Introduction}
\label{intro}
Data driven insights can improve corporate decision making. In large organizations, a variety of financial data is available due to reporting requirements and organizational purposes. However, expertise in the field of data analytics is scarce. New methods for robotic data evaluation can help organizations cut costs by shifting resources away from manual tasks and towards tasks that require supervision. Managers can derive valuable insights from forecasts based on internal company data to deal with common problems in a corporate environment such as demand forecasting for supply chain planning \citep{yelland_bayesian_2010}, call center arrival times \citep{taylor_comparison_2007, taylor_density_2011} and \cite{oreshkin_rate-based_2016}, sales forecasting \citep{kolsarici_correcting_2015}, or mergers \& acquistions (M\&A) forecasting \citep{very_can_2012}. However, the available datasets are usually subject to non-stationarity and structural breaks, and they usually make manual efforts to fit a meaningful model necessary. To deal with such problems, experts employ techniques on change point detection or finding stable parameter windows. To the best of our knowledge, no combination of such techniques was explored for a framework incorporating count data. To address this gap, we propose a method that automatically detects locally homogeneous time windows and corresponding parameters in an automated way that can be used to generate point or density forecasts.

\begin{figure}[ht]
    \centering
    \includegraphics[width=1\textwidth]{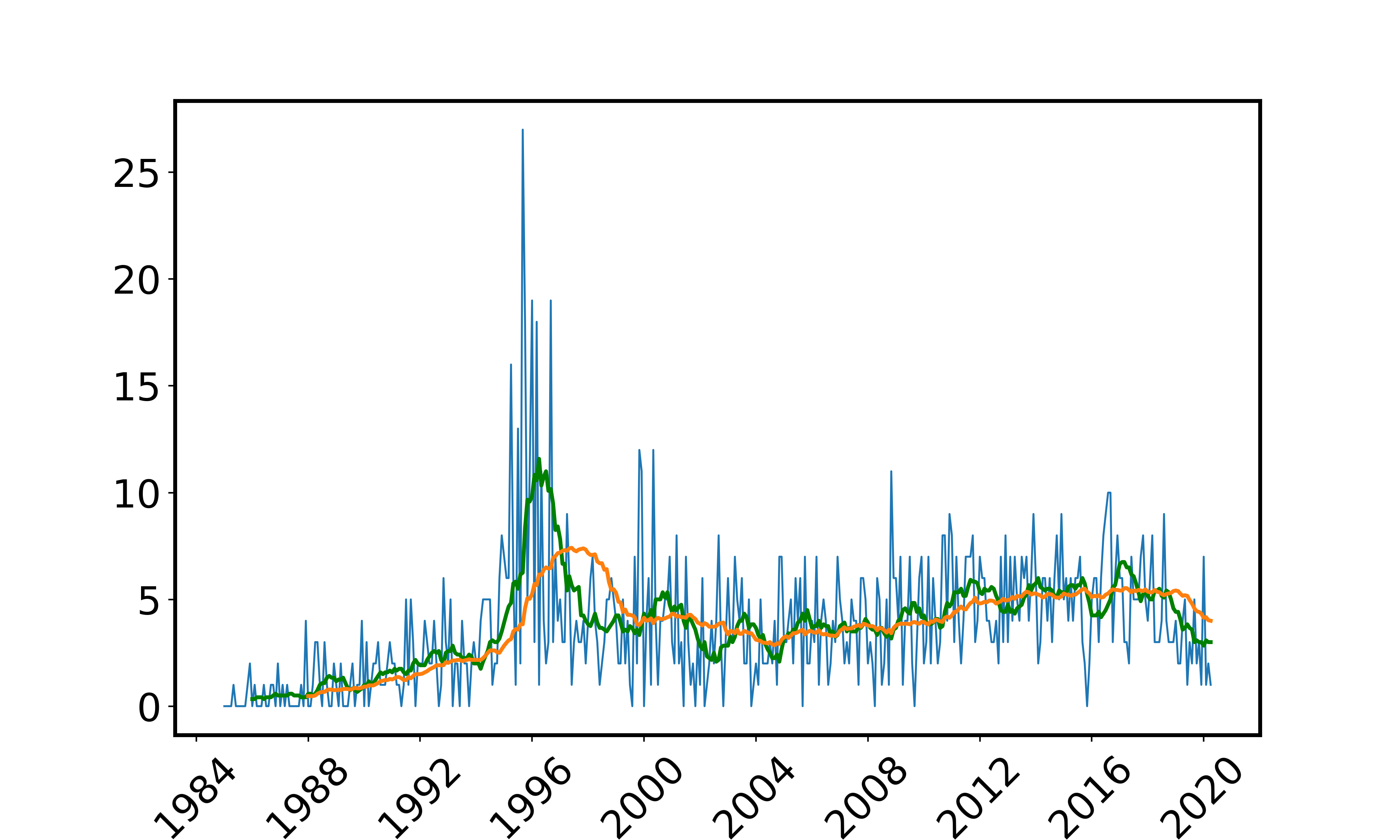}
    \caption{{\color{sapphirecrayola} Time series of count of mergers and acquisitions per month}, with moving average curve of {\color{aoenglish} 1 year} and {\color{orange}3 years}.
   \protect \includegraphics[height=0.5cm]{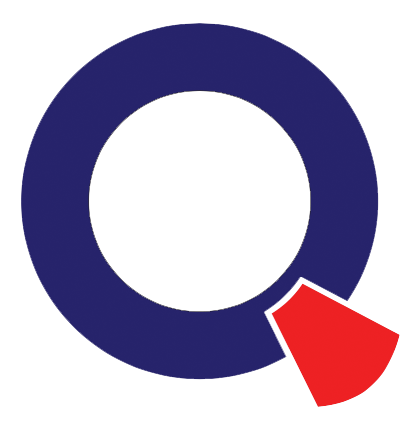} {\color{blue}\href{https://github.com/QuantLet/data_driven_controlling/tree/main/LPA_Empiricalstudy}{LPA\_Empiricalstudy}}
    }
    \label{fig:movingaverages}
\end{figure}

As a motivating example we apply this newly developed algorithm to forecast M\&A intensity in different industries. M\&A are especially interesting as they frequently occur in different markets and industries, and are relevant both to the financial industry that generates revenue by supporting their execution, as well as those companies that are observing their own industry and their competitors. Recent approaches are using time series models, as in \cite{very_can_2012} or a revealed preference model as in \cite{akkus_determinants_2015}. Figure \ref{fig:movingaverages} shows an example data set of mergers and acquisitions of German energy market that illustrates the presence of non-stationarity and structural breaks in such time series.

\sloppy

Empirical evidence strongly suggests that mergers are often clustered in time as waves, see \cite{martynova_century_2005},  \cite{harford_what_2005} and \cite{maksimovic_private_2013}. \cite{ahern_importance_2014} find that their activity is subject to network effects and that these waves largely occur within industries, but can also be transmitted to connected industries. Furthermore, shocks of any kind, even if they lead to merger waves, are difficult to predict. Wave patterns seem to be heterogeneous and differ both in time and in industries. Following their argumentation, we conclude that data on M\&A should be evaluated per industry and geographic location due to differences in regulation, innovation power, technology, and stock markets.

Predictive models for M\&A intensity could be identified through aggregating acquired knowledge and tailored to specific industries and markets. Alternatively, time-series models, e.g. ARMA can be used. While a co-variate based model provides explainability to the user, it requires manual efforts to gather data and incorporate expert knowledge to define relevant variables and calibrate their impact on a predictive model. Data gathering can be time-consuming and expensive, and knowledge about modeling heterogeneous industries in a specific application requires domain knowledge, which is often scarce and narrowed to said application. Time series models are an alternative, as they can be adapted to any other data set with comparable structure. However, such models need to be robust to non-stationarity, structural breaks and wave patterns that limit their predictive power. 

There is no doubt that only time varying time series models approximate the dynamics of the underlying series better than any homogeneous parameter approach, e.g. a fixed ARMA($p,q$) model. Therefore, we employ an adaptive estimation method called Local Parametric Approach (LPA). The quantitative implementation was first proposed in \cite{spokoiny_estimation_1998}, advances are made in \cite{mercurio_statistical_2004} and \cite{spokoiny_multiscale_2009}. It helps us to find locally homogeneous time intervals with stable parameters and guarantees a trade-off between parameter variance and modeling bias. The technique is based on a series of likelihood ratio tests to determine assumed but unknown change points in the underlying series. As a result one finds local intervals of homogeneity and efficient estimates at each point in time. 

Since we are often dealing with small sample sizes, and the test statistic distribution is unknown, we need a method to approximate this distribution. Recent advances in bootstrapping methodology allow us to generate confidence sets and critical values that non-asymptotically approximate the true distribution. Here, we couple LPA with multiplier bootstrap (MBS) \citep{spokoiny_bootstrap_2015} for approximating a critical value for the testing procedure. MBS builds up on wild bootstrap, that originates from \cite{wu_jackknife_1986} and \cite{beran_discussion_1986}. An important application is reported in \cite{hardle_comparing_1993}, and \cite{mammen_bootstrap_1993}. Further advancements are made in \cite{chatterjee_generalized_2005} and \cite{arlot_nonasymptotic_2010}. Notable publications that precede \cite{spokoiny_bootstrap_2015} are
\cite{bucher_multiplier_2013} and \cite{chernozhukov_gaussian_2013}. \cite{klochkov_localizing_2019} present an application in the context of a conditional autoregressive Value at Risk model. We generalize this LPA idea to any data that is Poisson jump distributed, although our simulation study indicates that these assumptions could be relaxed to the general membership to any of the exponential families. This indication is useful in bridging business requirements with the robustness of novel statistical methods through flexible automated estimations, since many problems in strategic management, e.g. forecasting M\&A intensity, are subject to data sets for which strict assumptions on underlying distribution, stationarity, or absence of structural breaks may not be fulfilled.

We detect locally homogeneous windows by computing a non-parametric likelihood ratio statistics. This approach is related to the branch of change point detection methods. \cite{chen_parametric_2011}, \cite{eckley_analysis_2011}, and \cite{aminikhanghahi_survey_2017} summarize and evaluate diverse methods of change point detection. Notable approaches are \cite{hinkley_inference_1970}, \citep{hsu_detecting_1979}, \citep{haccou_likelihood_1987}, and \citep{chen_change_1999} that propose change point detection methods for gamma, exponentially, and normally distributed data respectively. \cite{kutoyants_optimal_1999} propose an adaptive procedure as well as theoretical properties for Poisson distributed data. \cite{chen_parametric_2011} provides the null distribution of a likelihood ratio test for Poisson distributed random variables, but they do not evaluate the efficiency of the procedure on real data. 

Both change point detection and homogeneous window approaches can serve as determinator for an optimal forecasting window. Recent approaches for optimal window selection are \cite{giraitis_adaptive_2013},
\cite{pesaran_optimal_2013}, and \cite{inoue_rolling_2017}. They address parameter instability and frequent structural breaks and indicate that adaptive window selection is favorable over choosing fixed window sizes, such as a 1 year or 3 year moving average. Since we aim at developing a generally applicable method that is robust to different data characteristics, we need an adaptive window. Hence, we pursue a nonparametric approach that is independent of knowledge about or assumptions on the dataset, except that the values are generated by a Poisson process with smooth but time varying intensity over some unknown time window.

Our approach extends the previous literature as it serves as a generic toolbox that could easily be adapted to other applications and gives density forecasts that can (but do not have to) be adjusted by incorporating knowledge from industry experts and is adaptable to arbitrary frequencies. Although we show how to forecast the density of M\&A, our methodology could also extend research in other areas that typically use Poisson processes.

The remainder of this paper is structured as follows: 
Section 2 describes the algorithm that is based on a combination of LPA, MBS and put into action in an iterative procedure. Section 3 contains an experimental study. We describe the evaluation method, verify the robustness of the presented algorithm in simulation scenarios, apply it empirically on a dataset of mergers \& acquisitions and show how it can be used to generate forecasts. Section 4 presents key results, such as robustness to diverse data inputs and adaptability to other applications. Section 5 discusses limitations, such as in the evaluation approach or computational costs and suggests next steps like density forecasting, introducing a judgemental component and extending the test-statistic. All numerical algorithms can be found on {\color{blue}\href{www.quantlet.de}{Quantlet.de}}
\includegraphics[height=0.5cm]{images/qletlogo_tr.png}.
 
\section{Methodology}
\label{methodology}
\subsection{Basic Idea}
\label{problemstatement}
Planning processes in corporate environments are based on internal financial data of different kinds. Traditionally, these problems have been solved using diverse time series models. However, it is difficult to use them since real time series often are non stationary and have structural breaks. Automated analyses can be beneficial as they make modelling easier. To contribute to solving this problem, we focus on detecting locally homogeneous intervals with stable parameters. To be more specific, we focus on a count data model where the time varying intensity determines a Poisson process. Take again figure  \ref{fig:movingaverages} as an example. We aim to detect the years of '95 - '97 as a structural break and capture therefore the non stationary component (a slight uptrend is observable). 
We find locally homogeneous windows, verifying that the procedure is working and show how it can be used to obtain density forecasts.

\subsection{Stochastics}
\label{model}

Let $Y_t\in \mathbb {N}, t=0,...,T  $ be a count data time series such as the count of M\&A series, see in figure  \ref{fig:movingaverages}. Think of $Y_t \sim Poisson(\theta)$, where $\theta$ represents the rate or average number of occurrences in a fixed interval. Since we allow for time variation in our model, for any interval $I = [a, b]$ with $a<b$ and $a,b \in \{0, . . . , T\}$, we write $(Y_t)_{t\in I} \sim Poisson(\theta)$ 

The log likelihood function on $I$ is:
\begin{equation}
\label{eq:likelihood}
\begin{aligned}
L_I(\theta)&  = \sum_{t\in I}\log ( \theta^{Y_t}e^{-\theta}/Y_t!) 
         &= \log\theta\sum_{t\in I}Y_t -\sum_{t\in I}\theta-\sum_{t\in I}\log(Y_t!)
\end{aligned}
\end{equation}
the MLE  \({\tilde\theta_I}\) based on observations in \({i\in I}\) is:
\begin{equation}
\tilde\theta_I\eqdef \underset{\theta \in \Theta}{\text{argmax }} L_I(\theta)
\end{equation}
which for a Poisson model is the sample mean.

\subsection{Local Parametric Approach}
LPA, first introduced by \cite{spokoiny_estimation_1998} is based on the phenomenon that a series of locally parametric models can describe the features of a time series better than a global parametric model. The basic idea is that given a time series and a model for its dynamics, one finds locally stationary intervals of the time series in an online fashion. This is done by finding the set of most recent observations, such that the model parameters are approximately stable in that interval. This set of time points is called \textit{interval of homogeneity}. Employing the same procedure at each point in time, one locally estimates the parameter \citep{hardle_local_2015}. The merit of LPA is that it does not require an explicit expression of the law of the dynamics of the parameter, but only assumes that the parameter is constant on some unknown time interval in the past \citep{spokoiny_multiscale_2009}.

In order to check the homogeneity of an interval $I=[a,b]$, LPA looks for some break point $\tau \in (a,b)$ such that $A_\tau = [a,\tau)$ has one parameter and $B_\tau = [\tau, b]$ has another parameter. If at least one break point exists in the interval $I$, we conclude that the interval is non-homogeneous \citep{klochkov_localizing_2019}.

The testing hypotheses are therefore:
\begin{equation} \begin{aligned}
H_0(I) &: (Y_{t})_{t\in I} \sim Poisson(\theta^*_{I}), \theta^*_{I} \in \Theta,\\
&\text{         vs} \\ 
H_1(I) &:  (Y_{t})_{t\in A_\tau} \sim Poisson(\theta^*_{A_{\tau}}), \theta^*_{A_{\tau}} \in \Theta,\\
&(Y_{t})_{t\in B_\tau} \sim Poisson(\theta^*_{B_\tau}), \theta^*_{B_{\tau}} \in \Theta, \\  
&\text{with some } \theta^*_{A_{\tau}} \neq \theta^*_{B_{\tau}}  
\end{aligned}
\label{eq:hypothesis}
\end{equation}
The LR test statistic for a breakpoint $\tau$ is:
\begin{equation}
    \mathcal{T}_{I,\tau}= L_{A_{\tau}} (\tilde\theta_{A_{\tau}}) +  L_{B_{\tau}} (\tilde\theta_{B_{{\tau}}}) - L_{I} (\tilde\theta_{I})
\end{equation}
Since one has many candidates $\tau \in J$, one arrives at:
\begin{equation}
    \mathcal{T}_{I}=\max_{\tau\in J}\mathcal{T}_{I,\tau}
    \label{eq:test_stat}
\end{equation}

This unfortunately has a very intractable distribution, hence the critical values $\mathfrak{z}_{I}(\alpha)$, indicating that the test indicates that the test statistic rejects $H_0$ in equation (\ref{eq:hypothesis}) i.e. when
\begin{equation}
    \mathcal{T}_{I} \geq \mathfrak{z}_{I}(\alpha)
\end{equation}
is hard to calculate. Indeed the limiting distribution of $ \mathcal{T}_{I}$ is different from general likelihood ratio tests due to the presence of nuisance parameters (breakpoints) in the alternative hypothesis which are not identified under the null hypothesis. Hence, convergence of the generalized LR statistics to a $\chi^2 $ distribution according to Wilk's phenomenon can not be put into action. While the asymptotic distribution of the sup-LR test in equation (\ref{eq:test_stat}) can still be derived \citep{andrews_optimal_1994}, a large enough sample size is required for its asymptotic critical values to be applicable. Certainly, that is not the case in most of the practical situations where only small samples of data are available.

\cite{spokoiny_bootstrap_2015} provide a non-asymptotic result for mis-specified models with small sampling sizes. \sloppy The technique is called multiplier bootstrap (MBS) which is discussed in detail in the next section.

\subsection{Multiplier bootstrap}
\label{mbs}
Since the asymptotic distribution for the LR test statistic is not available for small samples, we approximate the unknown log-likelihood distribution using the bootstrap. First, introduce random weights to the previously defined likelihood function: 
\begin{ceqn}
\begin{align*}
   L_{I}^{\circ }(\theta) = \sum_{t\in I}w_tl_t(\theta)
\end{align*}
\end{ceqn}
where the weights \(w_t \text{ are with } \operatorname{E}(w_t)=1 \text{ and } \mathrm{Var}(w_t)=1 \text{ } iid\). The bootstrap version of (\ref{eq:likelihood}) is given by:
\begin{ceqn}
\begin{align}
   L_{I}^{\circ }(\theta) = \log \text{ } \theta \sum_{t\in I}Y_t w_t -\sum_{t\in I} w_t \theta-\sum_{t\in I} \log(Y_t!)w_t.
\end{align}
\end{ceqn}
The bootstrap MLE is then defined as:
\begin{ceqn}
\begin{align*}
   \tilde\theta_{I}^{\circ } = \arg \max L_{I}^{\circ }(\theta),
\end{align*}
\end{ceqn}
which is:
\begin{ceqn}
\begin{align*}
   \tilde\theta^{\circ }_{I} = \frac{\sum_{t \in I} Y_tw_t}{\sum_{t \in I} w_t}.
\end{align*}
\end{ceqn}
It follows that the corresponding bootstrap of (\ref{eq:T}) is:

\begin{equation}
\label{eq:T_bootstrap}
   \mathcal{T}_{I,\tau}^{\circ } ={} L_{A,\tau}^{\circ }(\tilde\theta_{A,\tau}^{\circ })+L_{B,\tau}^{\circ } (\tilde\theta_{B,\tau}^{\circ })
   -\sup_\theta\left \{  L_{A,\tau}^{\circ }(\theta)+ L_{B,\tau}^{\circ }(\theta+\tilde\theta_{B,\tau}-\tilde\theta_{A,\tau})\right \},
\end{equation}

A penalty term  \(\tilde\theta_{B,\tau}-\tilde\theta_{A,\tau}\) is introduced to compensate for model misspecification bias. \cite{klochkov_localizing_2019} shows that the distribution of this test conditional on the data mimics the "true" distribution of \( \mathcal{T}_{I}\) with high probability.
Using (\ref{eq:T_bootstrap}) we can obtain the critical value through simulations. Indeed, the critical value \(z_{I}^{\circ}(\alpha)\) is defined as: 
\begin{ceqn}
\begin{align}
   z_{I}^{\circ}(\alpha) = z_{I}^{\circ}(\alpha;Y) = \inf\left \{z\geq 0:\textup{P} ( \mathcal{T}_{I}^{\circ }> z^2/2)\leq \alpha \right \}.
\end{align}
\end{ceqn}

\subsection{Algorithm}
The algorithm for an adaptive window length selection at each point in time is now straightforward. It is based on sequential testing of the hypotheses on a nested set of intervals $\{I_k\}_{k=0,1, \ldots, K}$, where $I_0 \subset I_1 \subset \ldots \subset I_K $. Let $n_k = |I_k|$ be the number of observations in each interval. The first interval $I_0$ is assumed to be homogeneous with length $n_0$. Then, for each interval $I_k$, the null hypothesis of parameter homogeneity is tested against the alternative of a change point at an unknown location $\tau$ within $I_k$, as in (\ref{eq:hypothesis}).

\begin{figure}[h]
    \centering
    \includegraphics[width=1\textwidth]{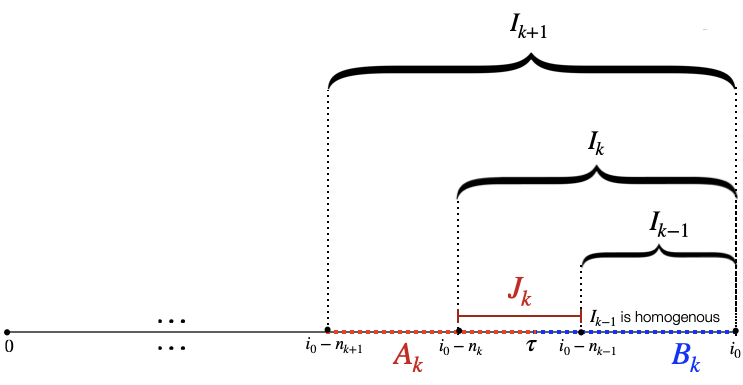}
    \caption{Iterative algorithm}
    \label{fig:algorithmplot}
\end{figure}

Since the setup deals with nested intervals, but the existence and location of a change point are unknown, only additional points in each new interval are considered as possible change points. The candidate set for change points in each interval is defined as $J_k = I_k \backslash I_{k-1}$. Using each point $\tau \in J_k$, the left and right intervals are constructed as $A_{k,\tau}=[i_0-n_{k+1}, \tau]$ and $B_{k,\tau}=(\tau, i_0]$ respectively (see figure (\ref{fig:algorithmplot})). The test statistic is calculated similarly to equation (\ref{eq:test_stat}) as
\begin{equation}
    \mathcal{T}_{I_{k},\tau}  = L_{A_{{k},\tau}} (\tilde\theta_{A_{{k},\tau}}) +  L_{B_{{k},\tau}} (\tilde\theta_{B_{{k},\tau}}) - L_{_{I_{k+1}}} (\tilde\theta_{I_{k+1}})
    \label{eq:T}
\end{equation}
where $A_{{k},\tau}$ and $B_{{k},\tau}$ are as previously specified and we test at every point $\tau \in J_k$ for a change point. The $k$th interval is rejected if
\begin{equation}
\max_{j \in J_k} \mathcal{T}_{I_{k},\tau}   \leq \mathfrak{z}_{I_k}^{\circ }(\alpha)
    \label{eq:maxT}
\end{equation}
 and $\mathfrak{z}_{I_k}^{\circ }$ is generated via multiplier bootstrap as explained in the previous section. If the interval $I$ is not rejected, i.e. there exists no change point and it is homogeneous, we continue the testing procedure by choosing a bigger interval. Otherwise, the length of the last non-rejected interval \(\widehat{I}\) is the interval of homogeneity and \(\widehat{\theta}_{i_0} = \widehat{\theta}_{\widehat{I}} \) is the respective adaptive estimate of \(\widehat{I}\).

Homogeneity testing for $I_k$ utilizes also part of observations of $I_{k+1}$. Hence, the pre-definition of intervals is crucial. Following that, the choice of interval lengths affects the test results, and therefore requires careful selection. We employ a geometric increase of intervals like \cite{hardle_local_2015} and \cite{klochkov_localizing_2019}. Based on the initial length $n_0$, the intervals lengths are defined by 
 \begin{equation}
     n_k = \ceil[\bigg]{n_0c^k}
     \label{eq:interval_len}
 \end{equation}
where $c$ is a geometric multiplier, chosen slightly above 1 to ensure a monotonic increase of interval lengths, but not by a big margin. Furthermore, instead of taking a constant number of intervals $K$ for testing as proposed by \cite{hardle_adaptive_2014} and \cite{klochkov_localizing_2019}, we select $K$ to be the smallest integer such that the whole time series is covered under a geometrically increasing length.

\label{algohärdle}

In summary, the LPA algorithm for  the adaptive choice of an interval of homogeneity and the corresponding MLE is given by the following iterative procedure:

\begin{enumerate}
\item\textbf{Initialization}: Select \(I_0, I_1, I_2,\) and define \(J_1=I_1\backslash I_0, (\forall \tau \in J_1),\)  \(A_{1,\tau}=[i_0-n_2, \tau] \), \(B_{1,\tau}=(\tau, i_0] \).
\item\textbf{Iteration}: For each iteration, select \(I_{k-1}, I_k, I_{k+1},\), 
\(J_k=I_k\backslash I_{k-1},(\forall \tau \in J_k)\) 
\(A_{k,\tau}=[i_0-n_{k+1}, \tau],\) \(B_{k,\tau}=(\tau, i_0] \).
\item\textbf{Testing homogeneity}:
Calculate test statistics in equation (\ref{eq:T}) and select critical value with multiplier bootstrap. Test hypothesis in equation (\ref{eq:hypothesis}) using equation (\ref{eq:maxT}).  
\item\textbf{Loop}: If \(I_k\) is accepted, take the next interval \(I_{k+1}\). Otherwise set \(\widehat{I}\) to the latest non rejected \(I_k\). 
\item\textbf{Adaptive estimator}:  Take  interval \(\widehat{I}\) as interval of homogeneity and \(\widehat{\theta}_{i_0} = \widehat{\theta}_{\widehat{I}} \) as adaptive estimate of \(\widehat{I}\).
Repeat the procedure for each point in time (different \(i_0\))
\end{enumerate}

\section{Experimental results}
\label{experiment}

The following lines seek to answer the question whether or not we can generate better forecasts by using the described adaptive methodology, and whether or not the methodology is robust with respect to previously unseen and thus unpredictable patterns. We compare one-step-ahead and multi-period point parameter forecasts of the proposed locally adaptive procedure to a baseline of one year and three years moving averages in a pseudo-out-of-sample approach. We acknowledge that this is a fairly simple approach. A more sophisticated evaluation approach would be to generate multi-period density forecasts that could be evaluated using a tailored loss-function as in \cite{diebold_evaluating_1998, diebold_comparing_2015}, and \cite{gonzalez-rivera_density_2014}. However, as this is beyond the scope of this paper (or altogether another paper), we leave it open for future work.

\subsection{Simulation study}
\label{simulation}

\begin{figure}[!htb]

\centering

\subfloat[]{
\includegraphics[width=.3\textwidth]{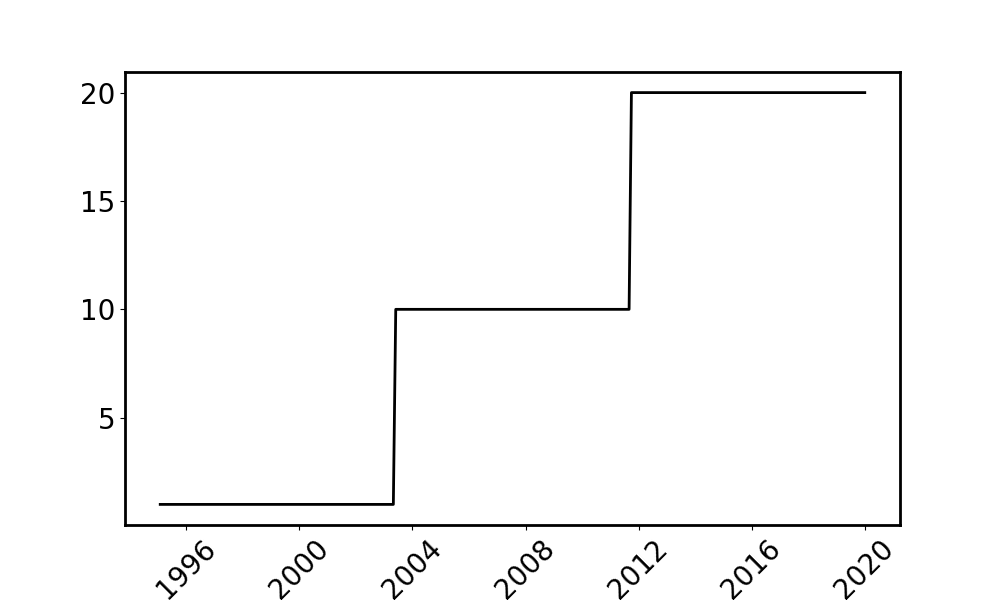}
\includegraphics[width=.3\textwidth]{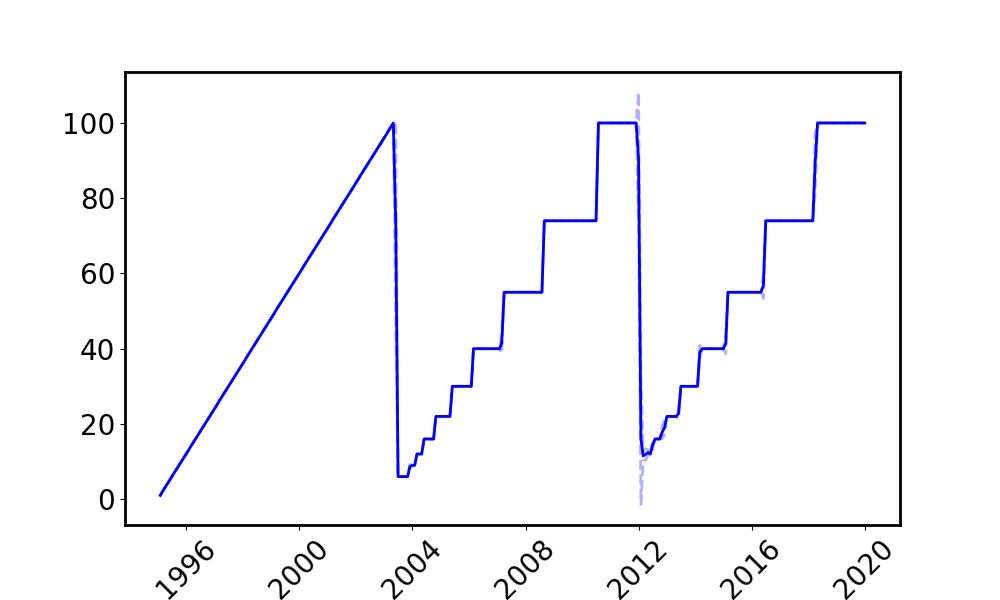}
\includegraphics[width=.3\textwidth]{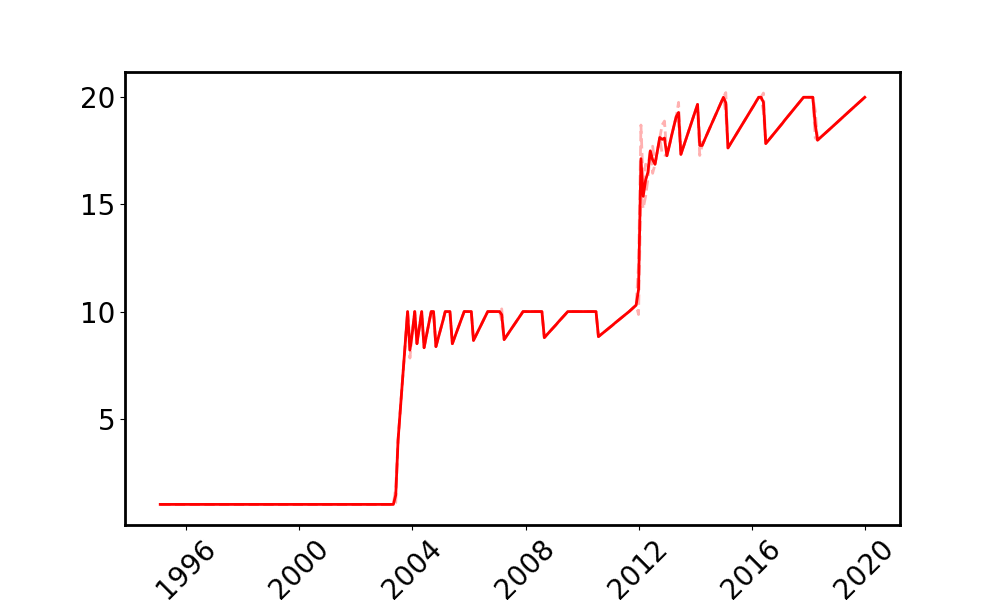}
}


\subfloat[]{
\includegraphics[width=.3\textwidth]{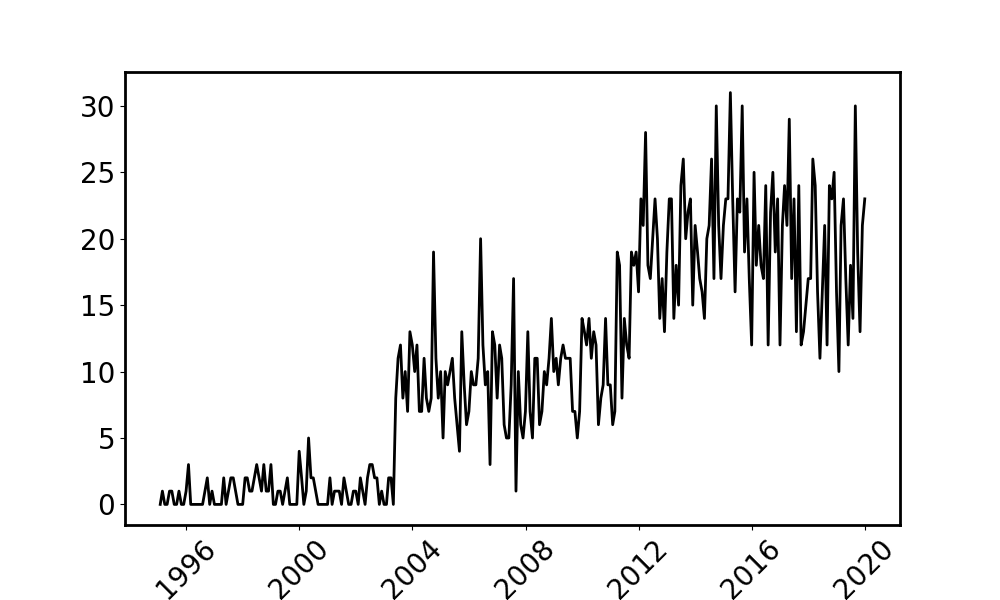}
\includegraphics[width=.3\textwidth]{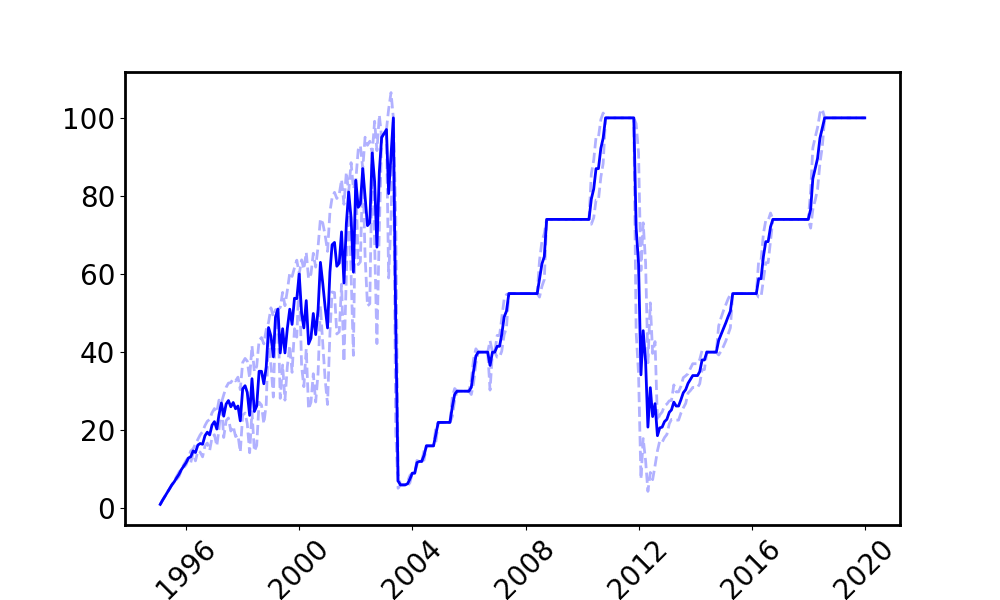}
\includegraphics[width=.3\textwidth]{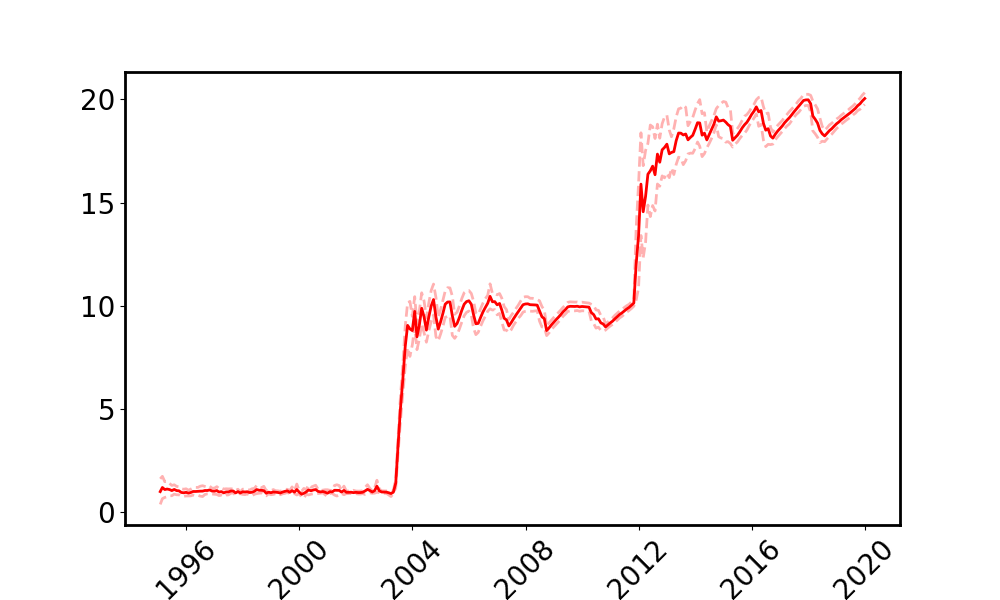}
}

\subfloat[]{
\includegraphics[width=.3\textwidth]{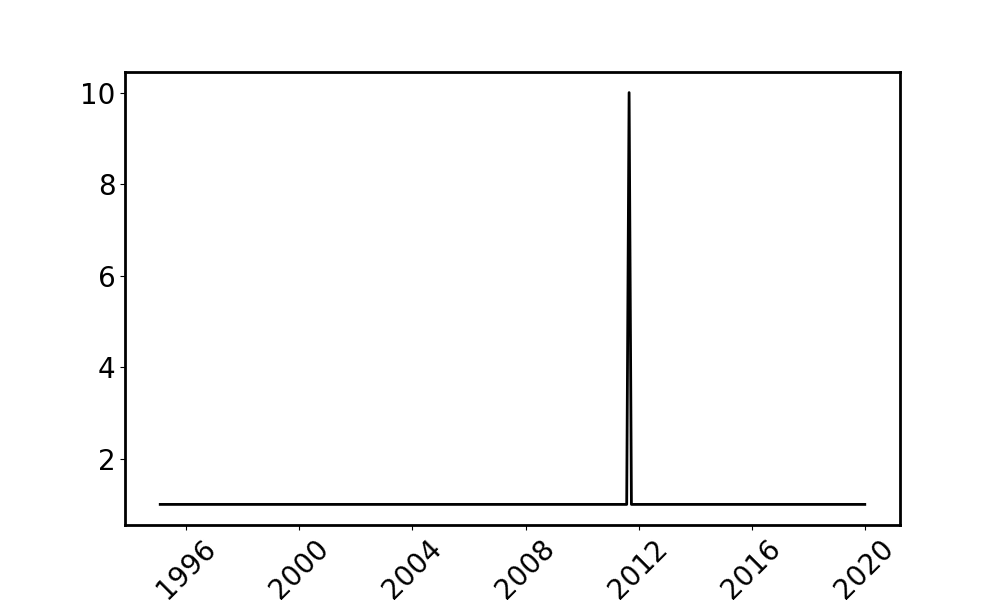}
\includegraphics[width=.3\textwidth]{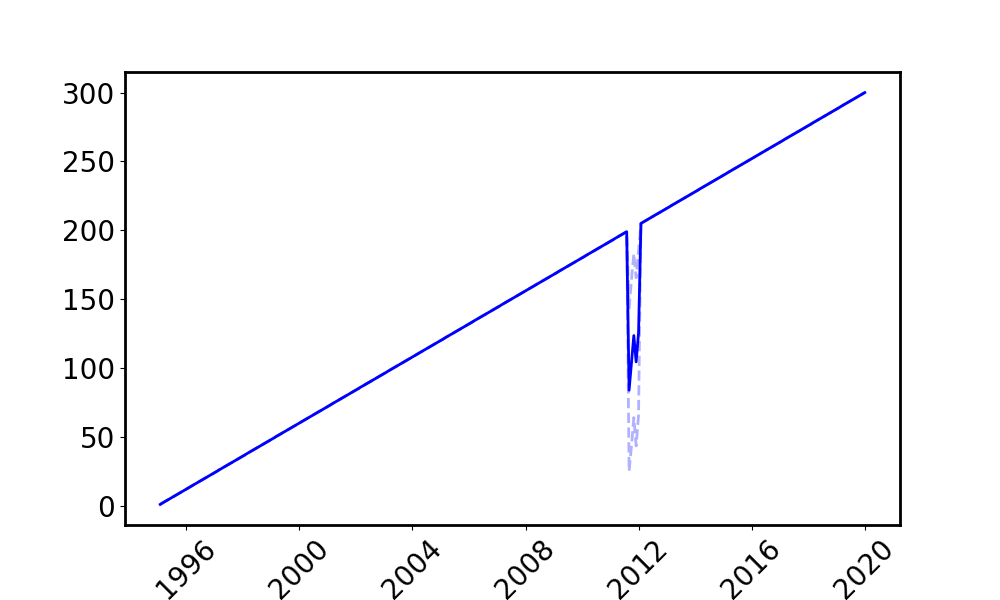}
\includegraphics[width=.3\textwidth]{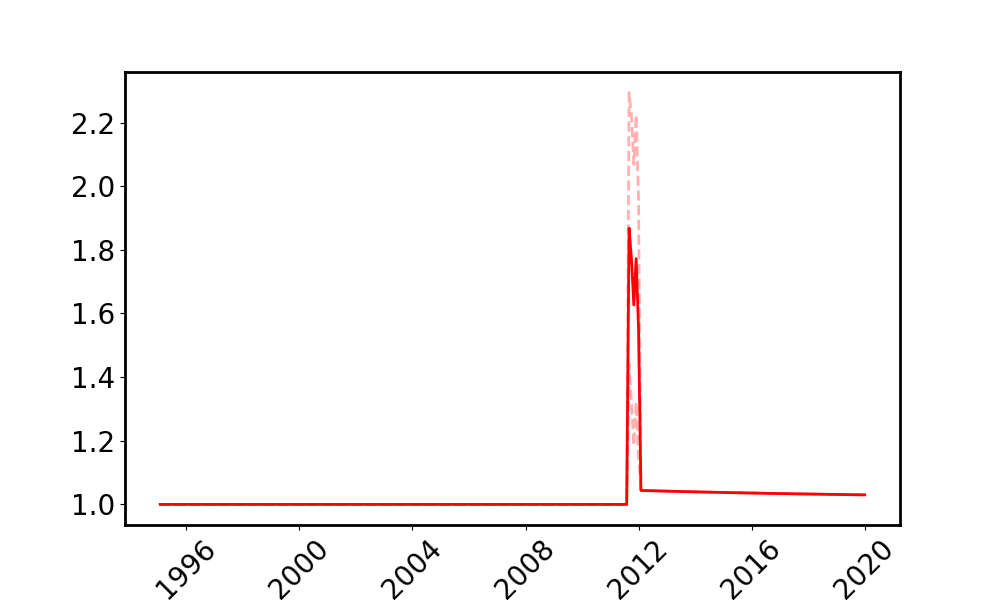}
}

\subfloat[]{
\includegraphics[width=.3\textwidth]{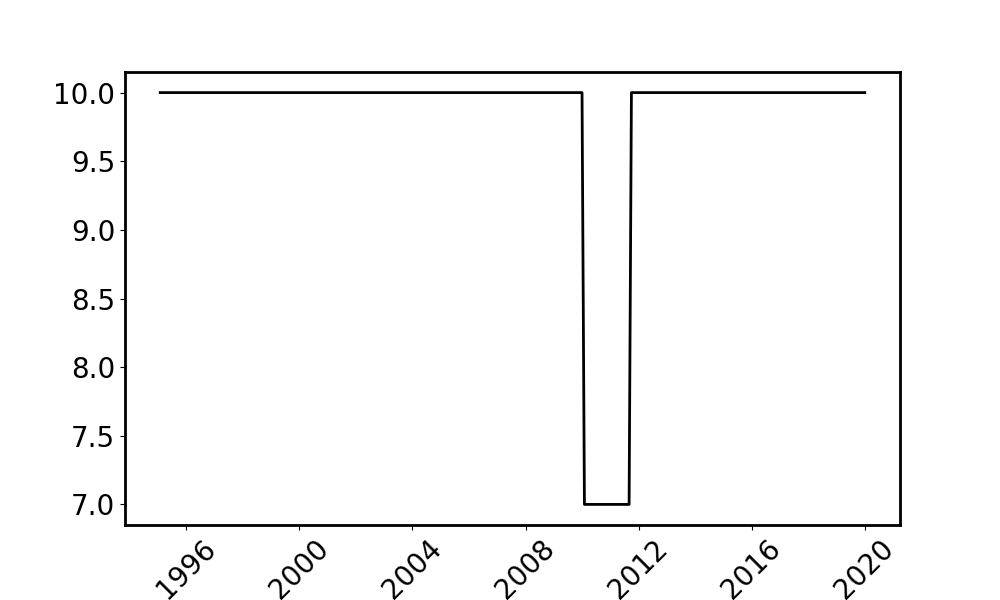}
\includegraphics[width=.3\textwidth]{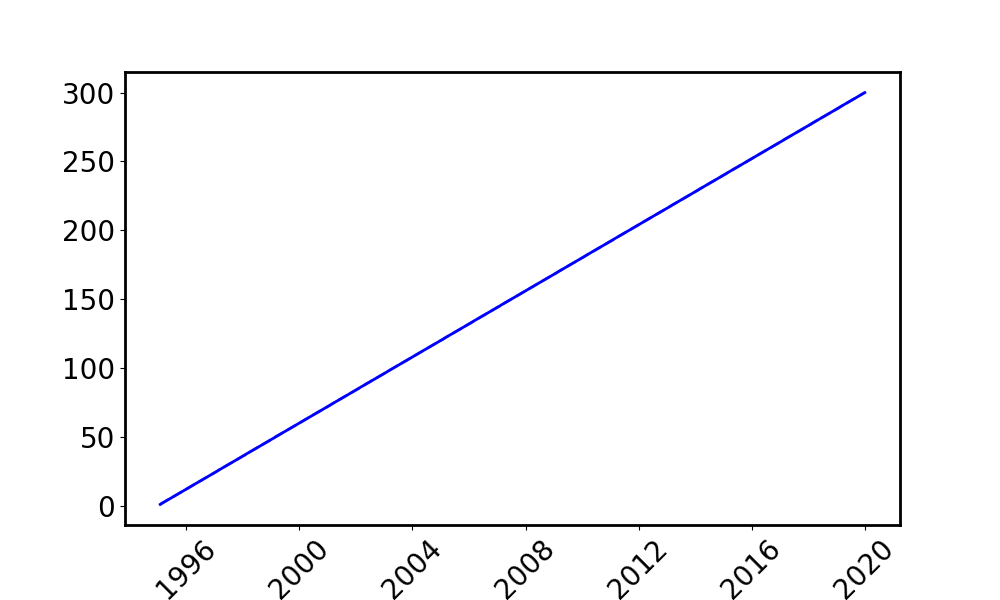}
\includegraphics[width=.3\textwidth]{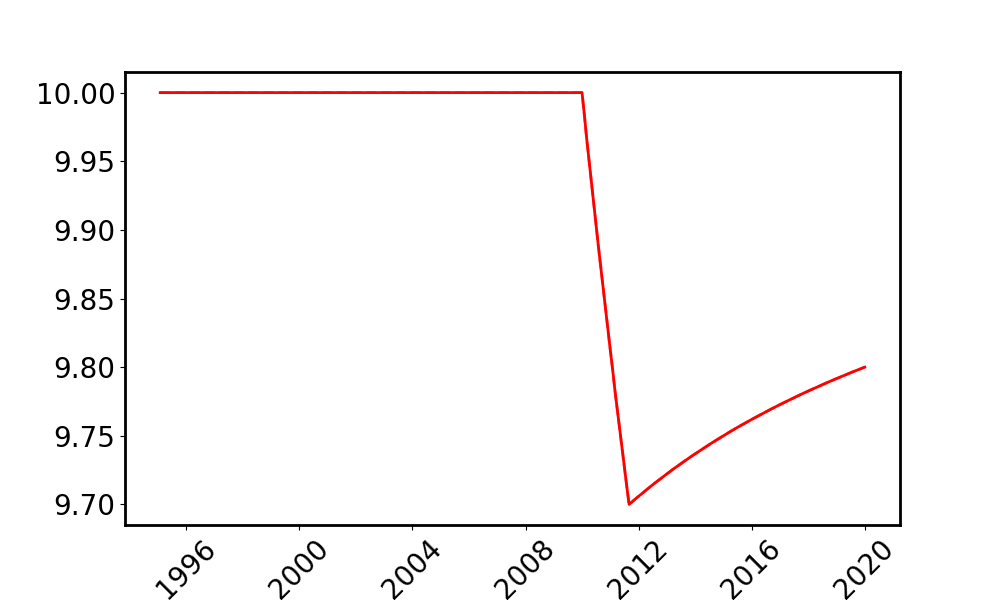}
}

\subfloat[]{
\includegraphics[width=.3\textwidth]{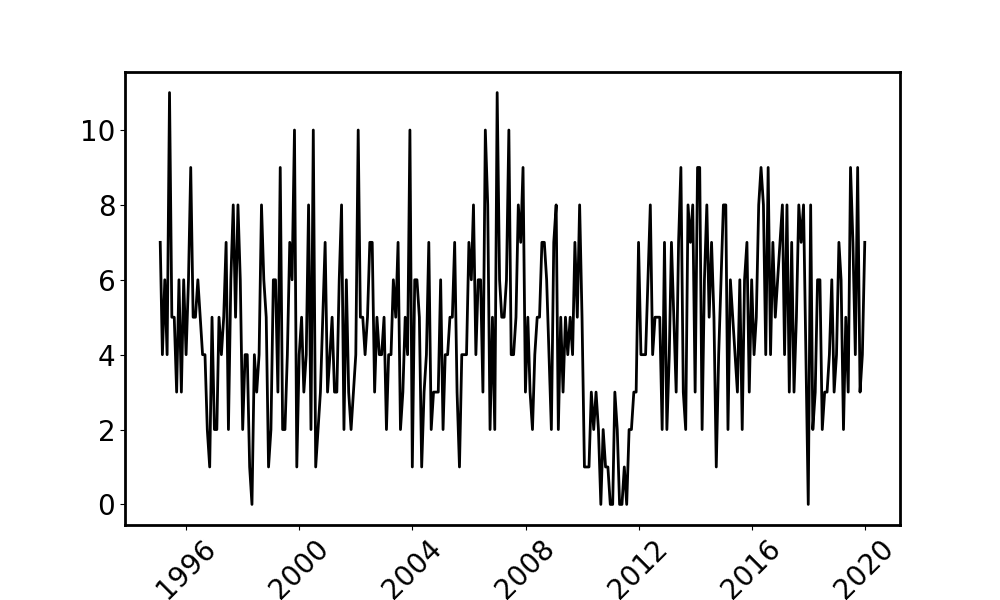}
\includegraphics[width=.3\textwidth]{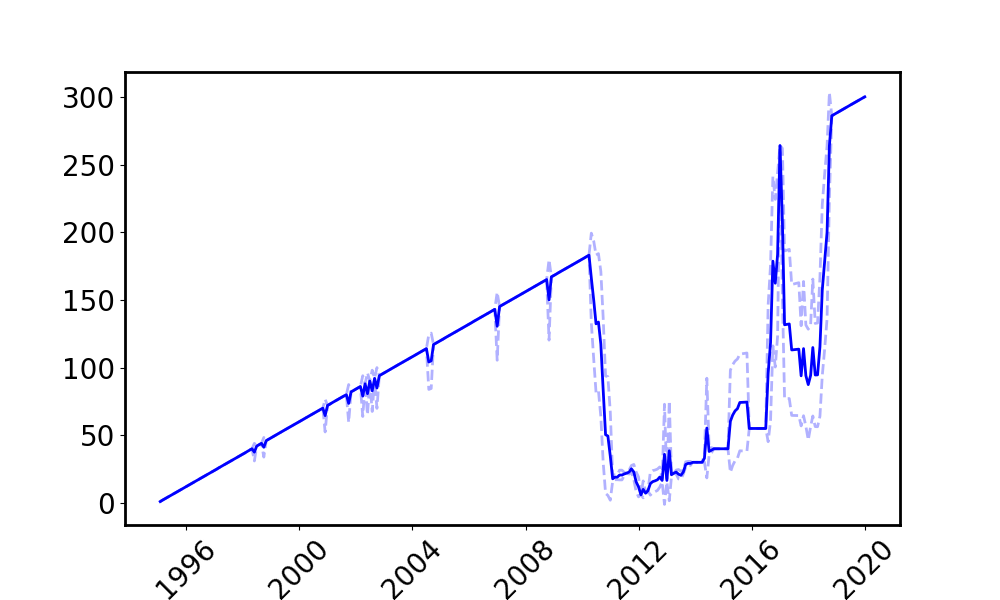}
\includegraphics[width=.3\textwidth]{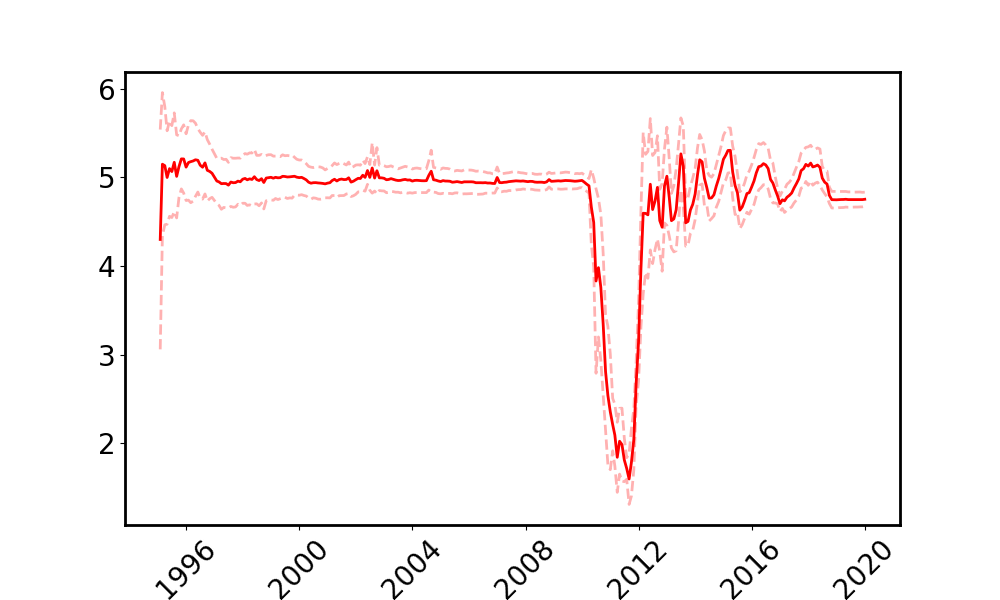}
}

\subfloat[]{
\includegraphics[width=.3\textwidth]{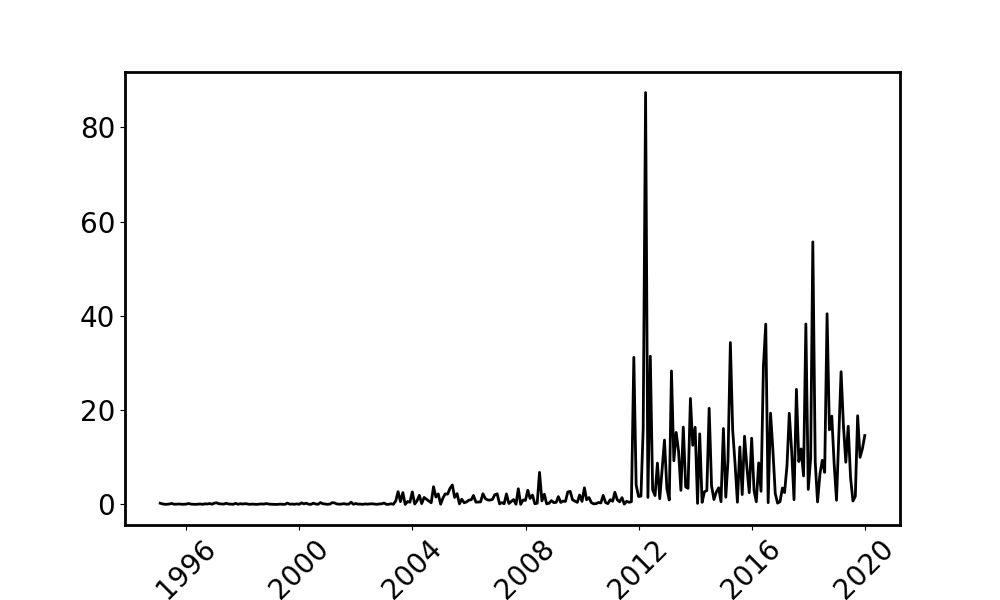}
\includegraphics[width=.3\textwidth]{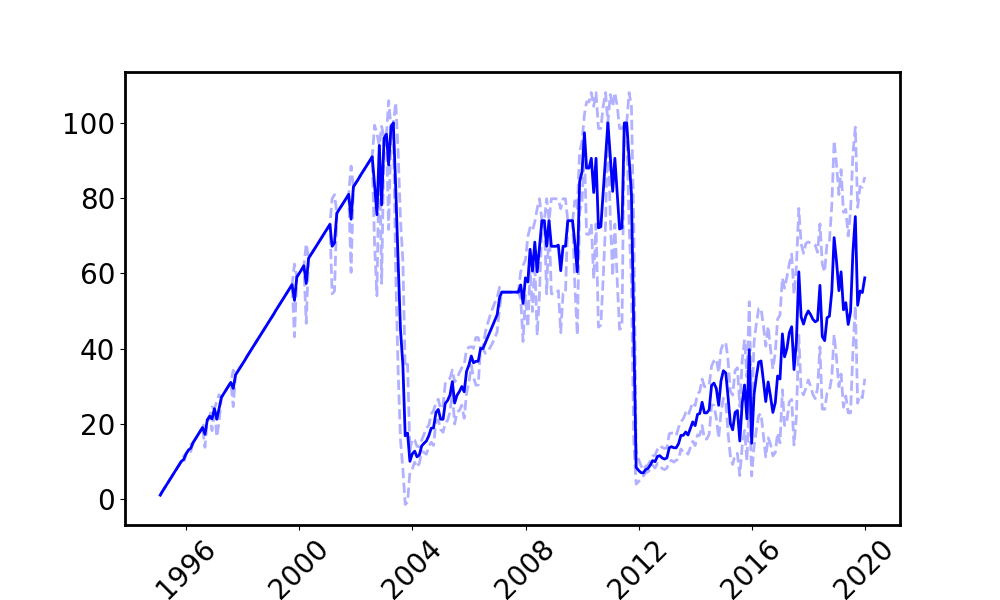}
\includegraphics[width=.3\textwidth]{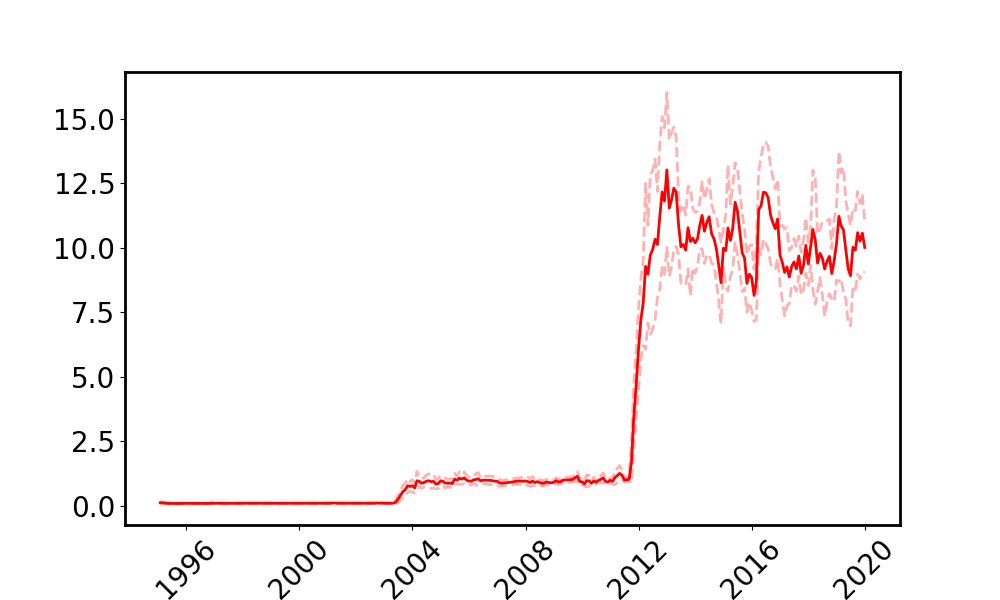}
}

\caption{Simulated series (left), homogeneous windows (middle) and MLE (right).
\centering{\protect \includegraphics[height=0.5cm]{images/qletlogo_tr.png} {\color{blue}\href{https://github.com/QuantLet/data_driven_controlling/tree/main/LPA_Simulations}{LPA\_Simulations}}}}
\label{fig:simulationplots1}
\end{figure}

This section evaluates the performance of the proposed technique using simulated datasets as shown in figure \ref{fig:simulationplots1}. We create scenarios that mimic common patterns in financial datasets. The simulations focus both on short-term shocks and regime shifts. Starting with the simplest piece-wise constant model, we gradually increase the complexity of the simulations by generating Poisson distributed data and finally test the robustness of the methodology by changing the underlying model to follow an exponential distribution. For each scenario, we consider a time series $(Y_{t})_{t=1}^{300}$ with the following specifications:

\begin{enumerate}[label=(\alph*)]
    \item \textit{Regime shifts with piece-wise constant model}: $(Y_{t})_{t=1}^{100}=1$, $(Y_{t})_{t=101}^{200} = 10$  and  $(Y_{t})_{t=201}^{300} = 20$
    \item \textit{Regime shifts with Poisson model}: $(Y_{1t})_{t=1}^{100}$ with $\theta_1=1$, $(Y_{2t})_{t=101}^{200}$ with $\theta_2=10$ and  $(Y_{3t})_{t=201}^{300}$ with $\theta_3=20$
    \item \textit{Short term shock with piece-wise constant model}: $(Y_{t})_{t=1}^{199}=1$, $(Y_{t})_{t=200}^{200} = 10$  and  $(Y_{t})_{t=201}^{300} = 1$
    \item \textit{Structural break with piece-wise constant model}: $(Y_{t})_{t=1}^{180}=10$, $(Y_{t})_{t=181}^{200} = 7$  and  $(Y_{t})_{t=201}^{300} = 10$
    \item \textit{Structural break with Poisson model}: $(Y_{1t})_{t=1}^{180}$ with $\theta_1=5$, $(Y_{2t})_{t=181}^{200}$ with $\theta_2=1$ and  $(Y_{3t})_{t=201}^{300}$ with $\theta_3=5$
    \item \textit{Regime shifts with exponential model}: $(Y_{1t})_{t=1}^{100}$ with $\theta_1=0.1$, $(Y_{2t})_{t=101}^{200}$ with $\theta_2=1$ and  $(Y_{3t})_{t=201}^{300}$ with $\theta_3=10$
\end{enumerate}

\FloatBarrier


The simulated data of scenarios (a)-(f) are shown in the time series plots on the left hand side in figure (\ref{fig:simulationplots1}). Since the algorithm requires pre-selection of the intervals, we fix c in equation (\ref{eq:interval_len}) at 1.35. This seemingly arbitrary choice ensures that we neglect only few unknown homogeneous intervals (if any), while gaining computational efficiency. Assuming a minimal homogeneous window ($n_0$) of 5 months, we compute $K$ as described in section (\ref{algohärdle}). For example, the candidate homogeneous windows for the latest period in the time series are $[6, 9, 12, 16, 22, 30, 40, 55, 74, 100, 135, 183, 247, 300]$. Note that since we allow the number of intervals to vary with the time point for which the homogeneous windows has to be determined, the number of intervals being tested gradually decrease as the number of historical data points in the time series decrease. The algorithm is applied to find the homogeneous window and corresponding MLE estimate for each point in time in each simulation and the results are presented in figure (\ref{fig:simulationplots1}). Also, for each scenario, we run 100 simulations to smoothen the estimated window sizes and construct confidence intervals that are depicted by dotted lines in all plots in figure (\ref{fig:simulationplots1}). 

\begin{figure}[!htb]
\centering
\subfloat[]{
\includegraphics[width=.3\textwidth]{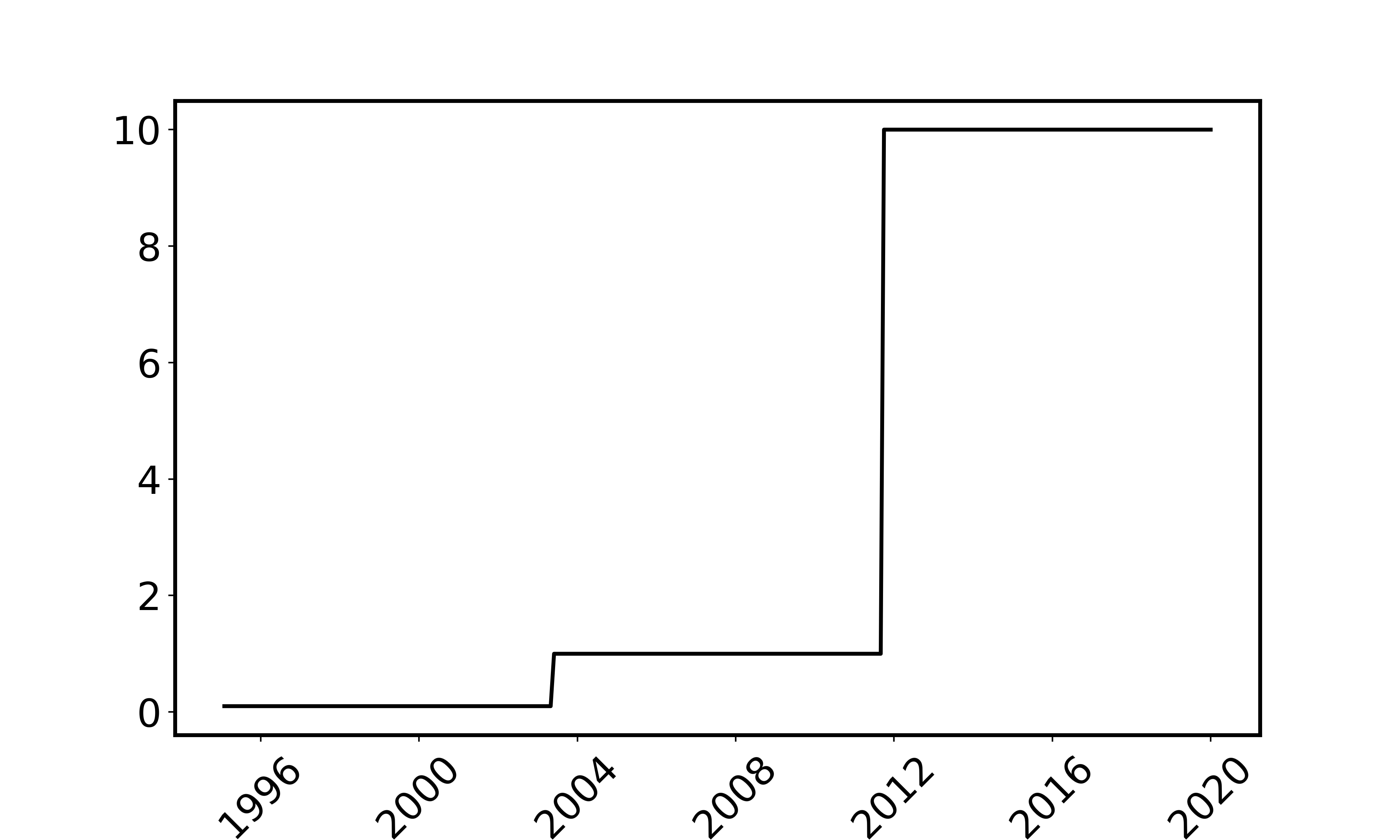}
\includegraphics[width=.3\textwidth]{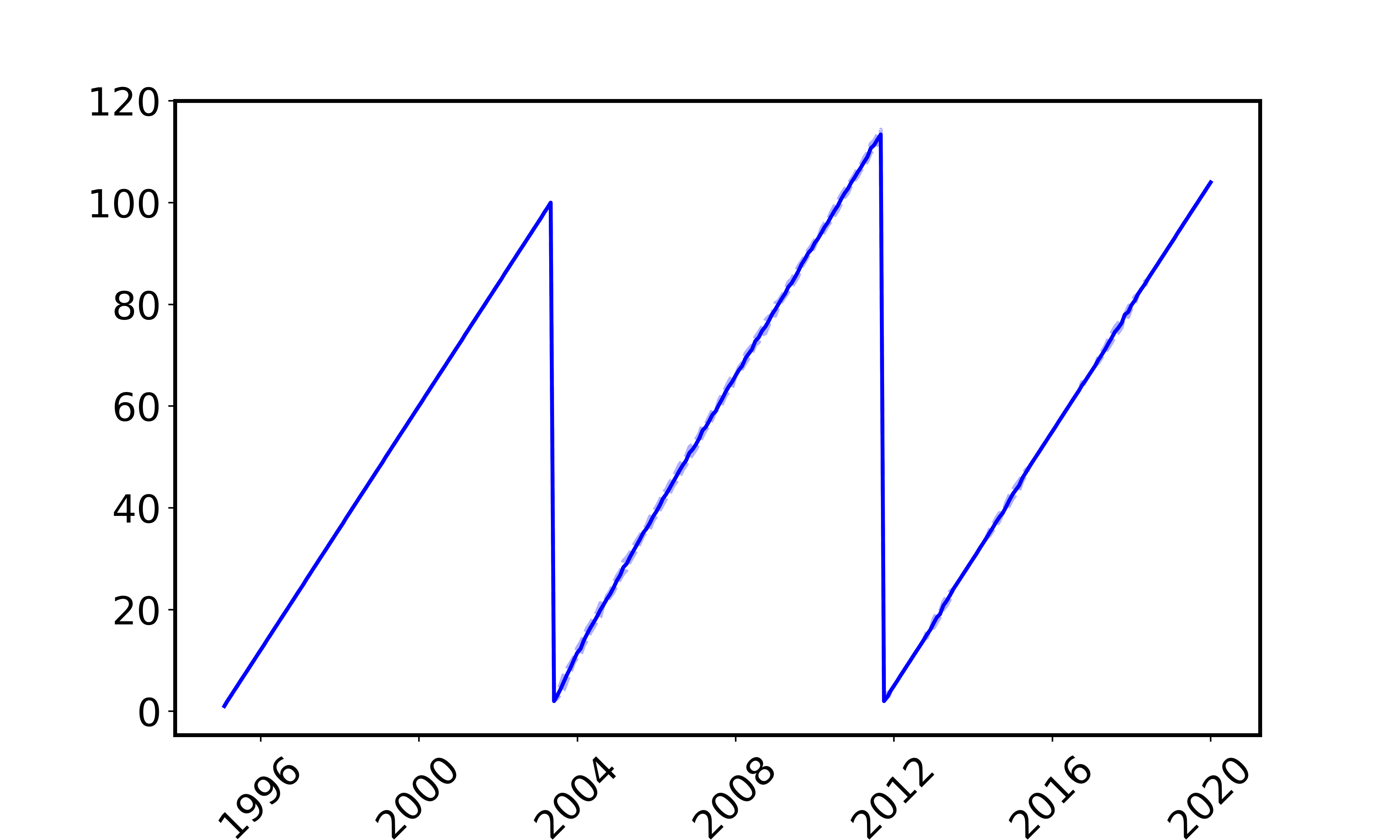}
\includegraphics[width=.3\textwidth]{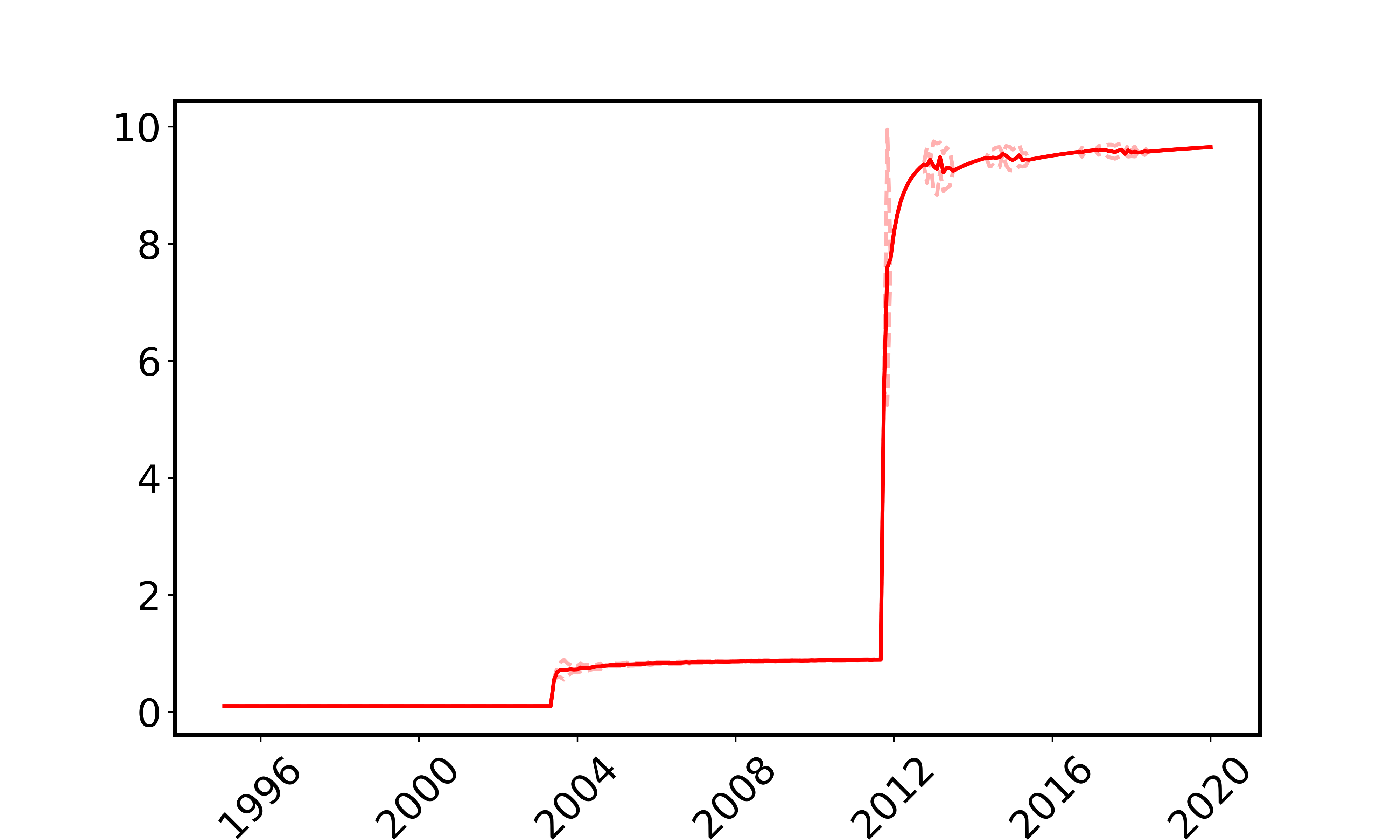}
}
\caption{Simulated series (left), homogeneous windows (middle) and MLE (right) using arithmetically increasing intervals in LPA. \protect \includegraphics[height=0.5cm]{images/qletlogo_tr.png} {\color{blue}\href{https://github.com/QuantLet/data_driven_controlling/tree/main/LPA_Simulations}{LPA\_Simulations}}}
\label{fig:simulationplots2}
\end{figure}

Plot (a) in figure (\ref{fig:simulationplots1}) shows that the procedure detects intervals of homogeneity correctly in all cases of this fairly simple scenario. It is robust to small and big values and gives a good approximation of the true parameters. The step-wise increase of homogeneous windows and the fluctuation of MLEs is a consequence of the limited number of possible time windows $K$ that the algorithm tests for. If $K$ were to be increased, for example arithmetically, we would expect the algorithm to generate a straight downward slopping line instead. Due to computational limitations, we illustrate this only on scenario (a) in figure (\ref{fig:simulationplots2}) and avoid such an experiment for other scenarios. 

\FloatBarrier

Next, a more realistic scenario with values generated from a Poisson process in (b) shows that the procedure is robust to noise even when the sample size is small. On the other hand, when $n$ is small, the problem of multiple testing causes many false negatives. This becomes evident in the first third of (b). This can be easily dealt with by changing the number of tested intervals in the algorithm. Regardless of this issue, the MLE remains close to the true parameter value. 

Further, some temporary shocks and small structural breaks are mimicked in simulations (c)-(e). (c) shows that a large shock is detected accurately, with an alarm to select smaller window size when the time is close to the change in mean. The simulation in (d) shows that temporary changes in mean are not always recognised by the algorithm, but the MLE after such short-term shocks is still affected. The procedure also detects structural breaks within a Poisson framework accurately as depicted in (e), but the estimated time windows are inaccurate shortly after the break. With increasing $n$, the accuracy is restored and the approximated MLE remains accurate. 

Finally, to check the robustness of the algorithm, values are generated from an exponential distribution. The simulation in (f) shows that the algorithm detects breakpoints correctly even with exponentially distributed data. Accordingly, the assumption of Poisson distributed values can be easily relaxed towards models from any exponential family. This robustness allows the algorithm to be used even with misspecified models. 

\begin{figure}[!htb]
\centering
\subfloat[]{
\includegraphics[width=.45\textwidth]{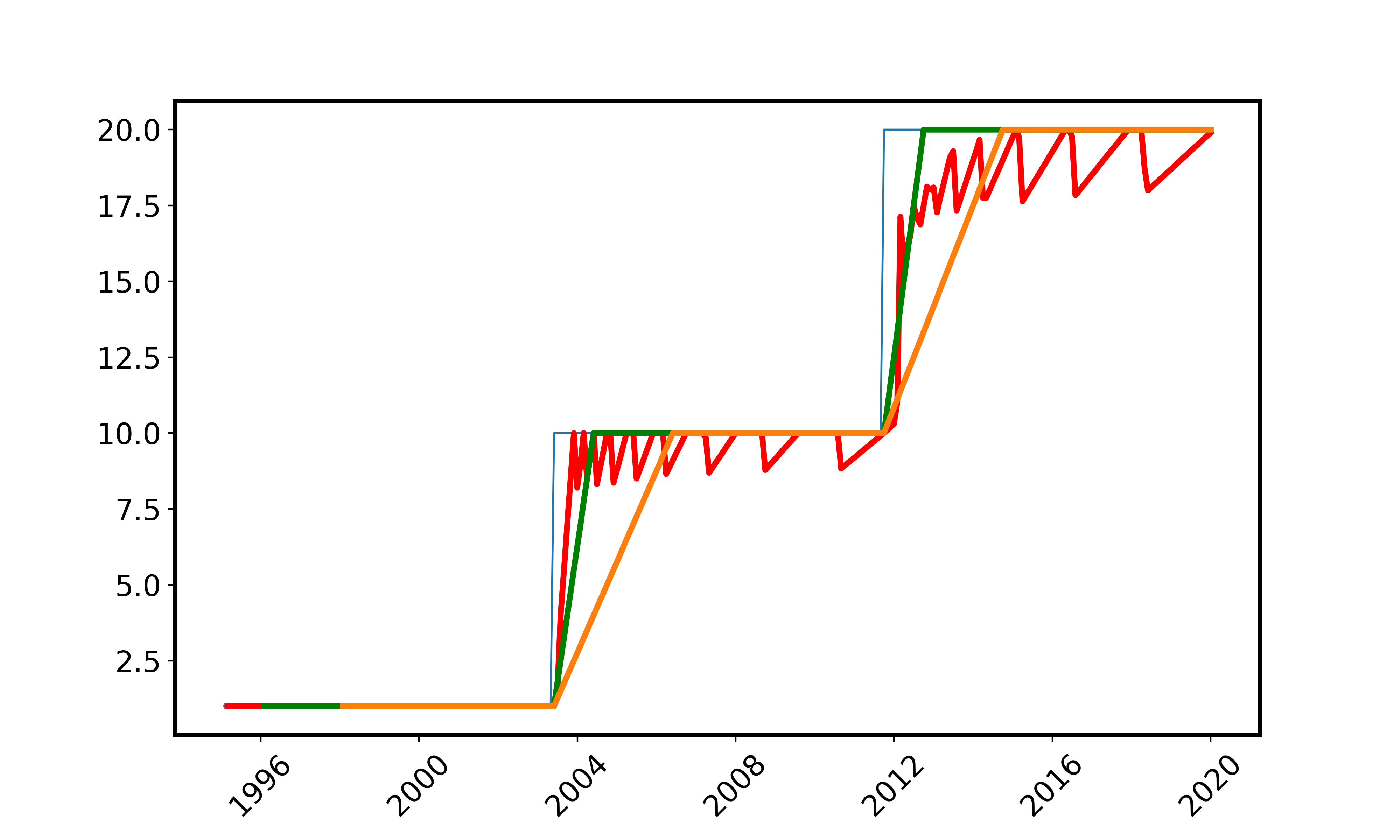} 
}
\subfloat[]{
\includegraphics[width=.45\textwidth]{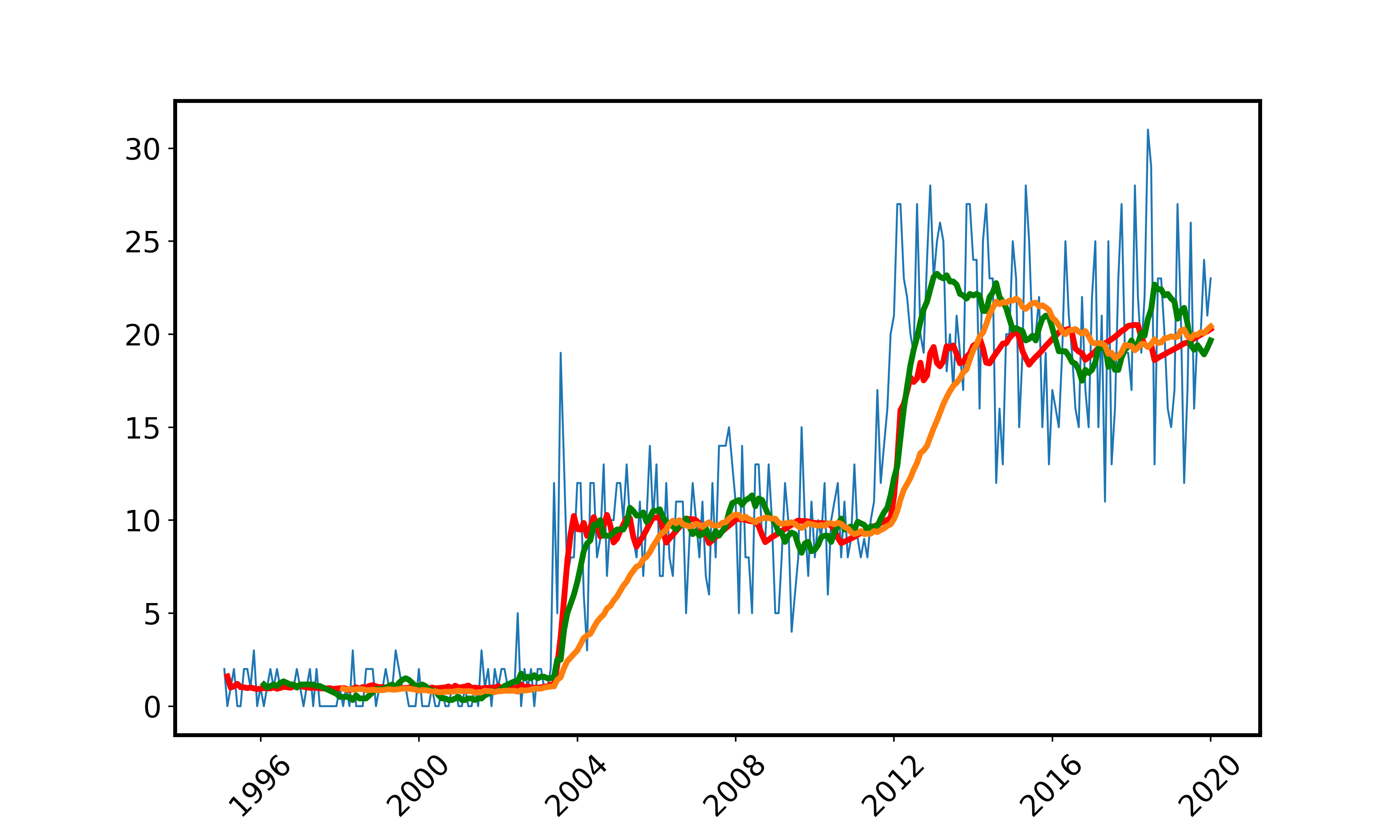} 
}

\subfloat[]{
\includegraphics[width=.45\textwidth]{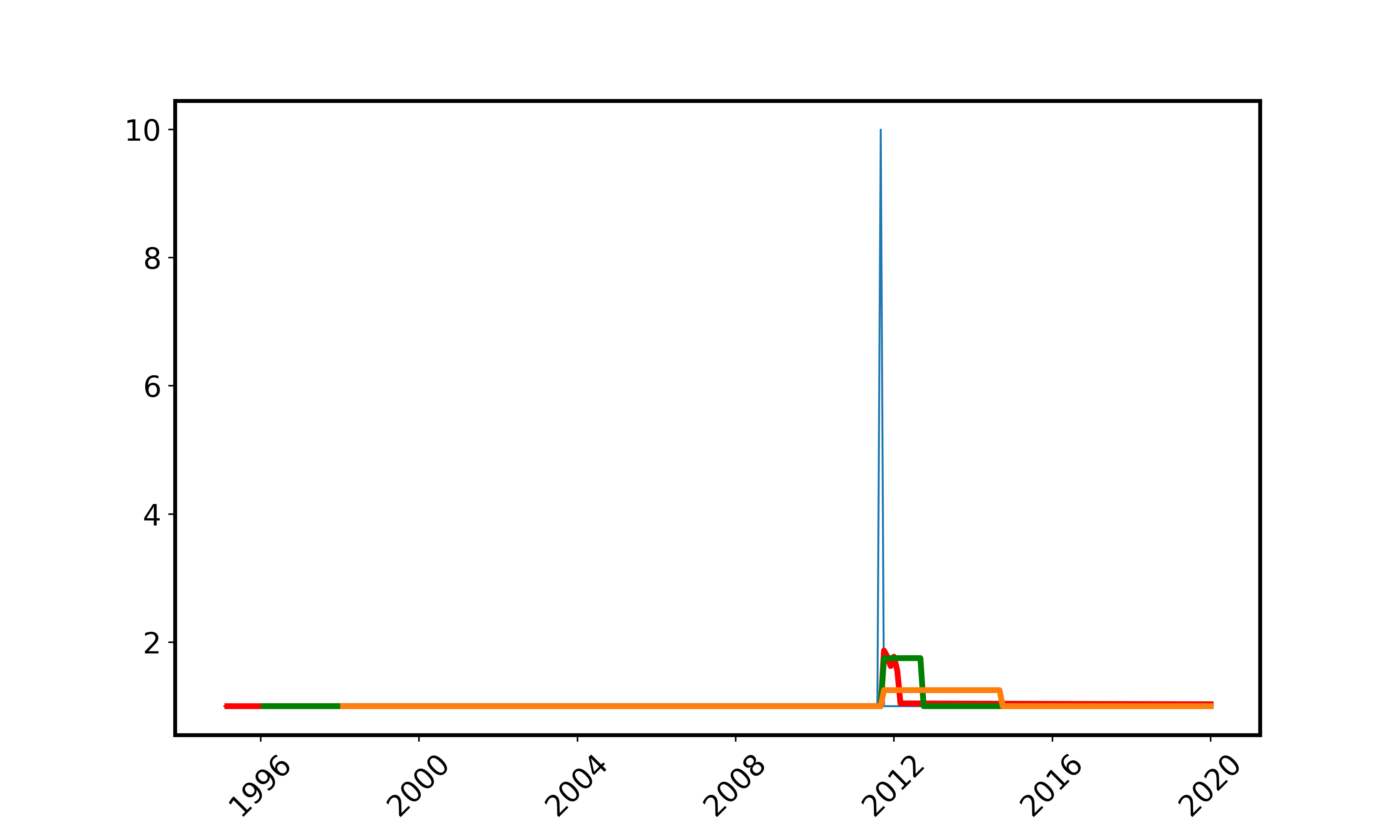}
}
\subfloat[]{
\includegraphics[width=.45\textwidth]{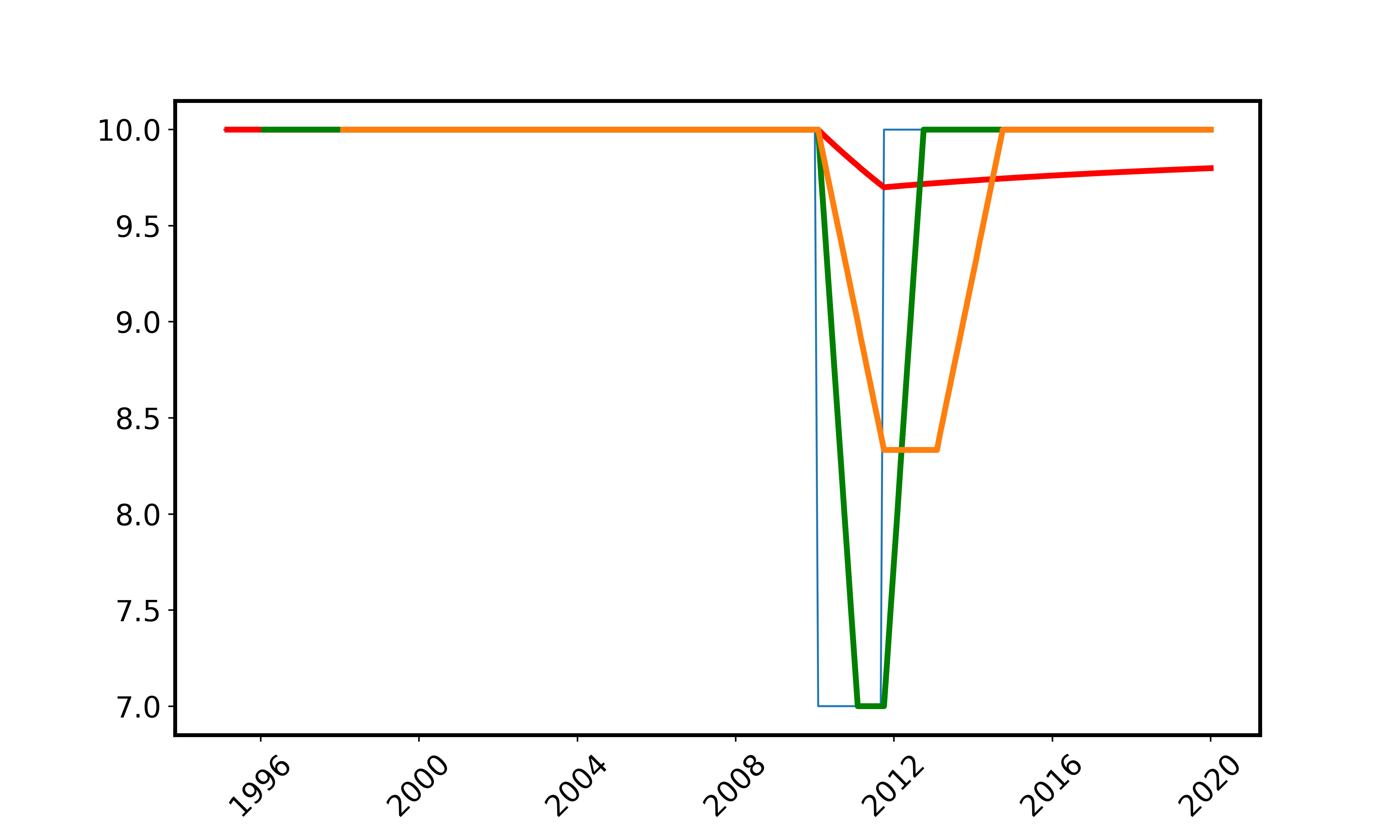}
}

\subfloat[]{
\includegraphics[width=.45\textwidth]{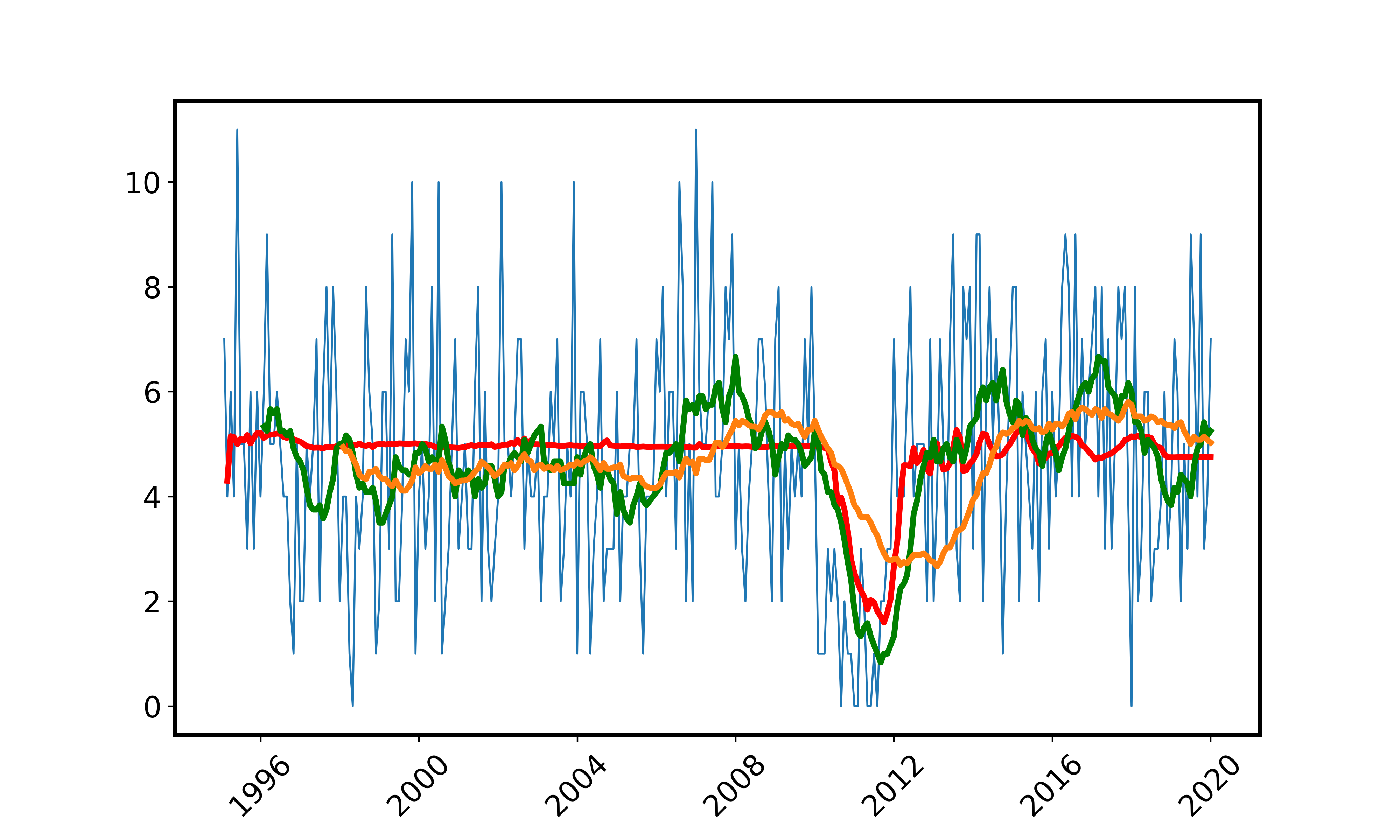}
}
\subfloat[]{
\includegraphics[width=.45\textwidth]{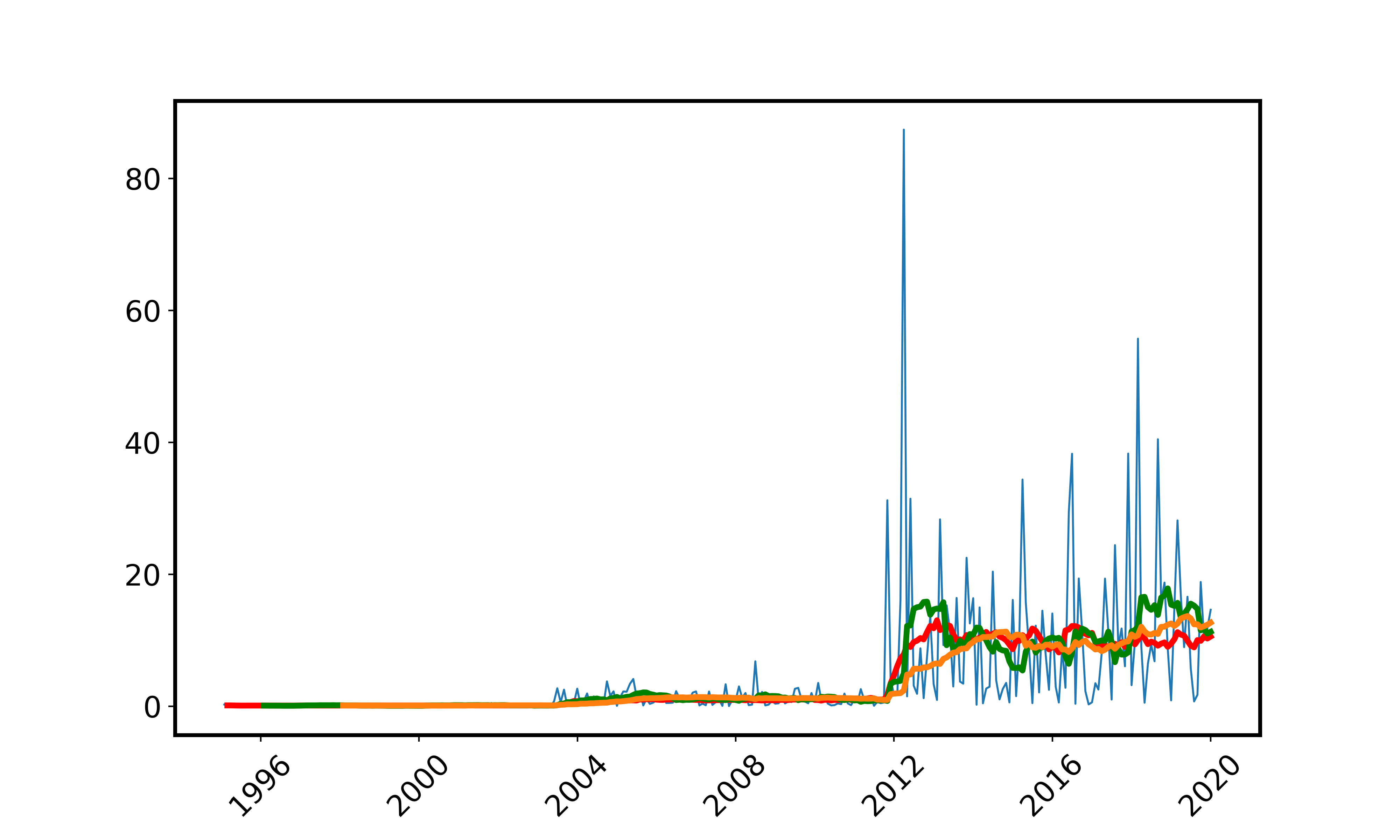}
}

\caption{{\color{sapphirecrayola} Time series of simulated data} and one step ahead prediction with estimates from {\color{red} LPA}, {\color{aoenglish}1 year fixed window} and {\color{orange} 3 year fixed window}. \\
\centering{ \protect \includegraphics[height=0.5cm]{images/qletlogo_tr.png} {\color{blue}\href{https://github.com/QuantLet/data_driven_controlling/tree/main/LPA_Simulations}{LPA\_Simulations}}}}
\label{fig:simulationplots_comparision}
\end{figure}

For each simulation scenario, we also compare the results with fixed window estimates of 12 months (high variance) and 36 months (high bias) and show the results in figure (\ref{fig:simulationplots_comparision}). All subplots show that LPA finds a balance between variance and bias by selecting a smaller or bigger window size when necessary. Plot (a) shows that while the estimates of LPA fluctuate (as described above), the shift in regime is detected earlier by LPA than by fixed window estimates. Plots (b) and (e) show that for the simulated intervals with constant mean, fixed window estimates indicate a highly fluctuating mean, while LPA recognizes intervals of time homogeneity correctly. Moreover, the impact of the short term shock disappears faster in the LPA estimates in (c) and (d). However, LPA underestimates the magnitude of shocks as visible in (d). 

In conclusion, the algorithm accurately detects small shocks, regime shifts and temporary mean changes in time series without strict assumptions on the underlying model. The results could be improved further by correct selection of interval lengths and numbers of intervals being tested, as they have a significant impact on the precision and accuracy of the results.  

\FloatBarrier

\subsection{Use case study}
\label{usecase_study}
In this section, we apply LPA to a working example of time series that are relevant to businesses in the financial industry. We use a data set of mergers \& acquisitions per month that we acquired from the database Refinitiv Eikon Deals (restricted access through subscription). We consider different industries according to the Thomson Reuters Business Classification scheme in Germany. The dataset consists of a total of 9,969 observations in ten industries in the US and Germany. We put the procedure into action on the following three industries in Germany:
 
\begin{enumerate}
    \item Financials
    \item Telecommunication
    \item Energy
\end{enumerate}

Due to seemingly inconsistent recordings of transactions, we considered only 424 observations from 01-1984 to 04-2020 in Germany for each industry. We do not make any assumptions regarding correlations of these industries or between countries.  

Following previous literature, we consider different industries separately. Without insider knowledge, the frequency of M\&As could be assumed to be random and i.i.d. over certain time windows, which is based on the intuition that we usually learn about mergers only after they are announced. The i.i.d. assumption states in this context that the observed industries are of a specific, somewhat stable structure in the short run, but can be interrupted by exogenous factors that change industry dynamics. Given these characteristics, we model the number of M\&A with a Poisson process and apply LPA to forecast the next period given the MLE of the last observed homogeneous time window. We do not preprocess the data and let the procedure account for structural breaks and non-stationarity. 

\begin{figure}[!htb]
\centering

\subfloat[]{
\includegraphics[width=.5\textwidth]{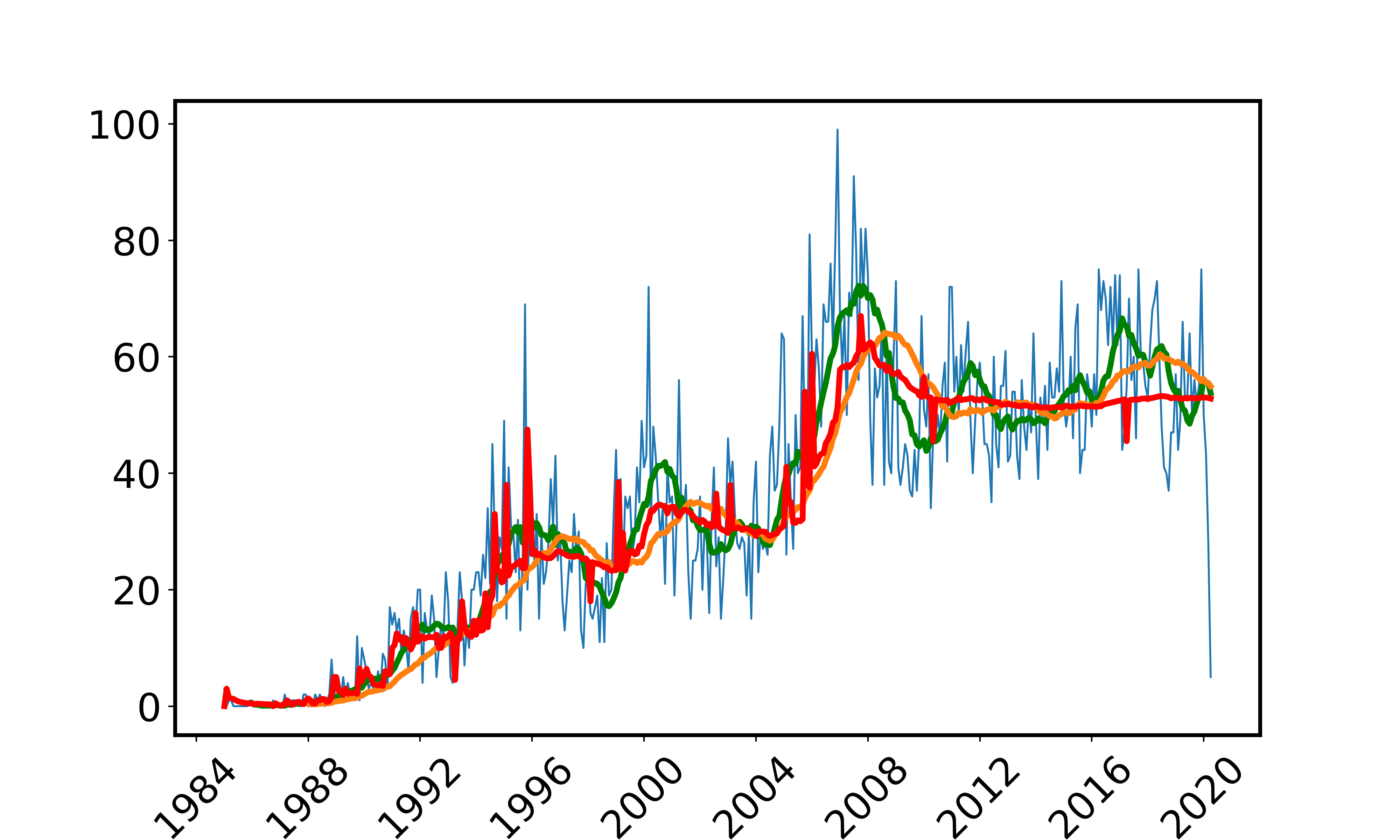}
\includegraphics[width=.5\textwidth]{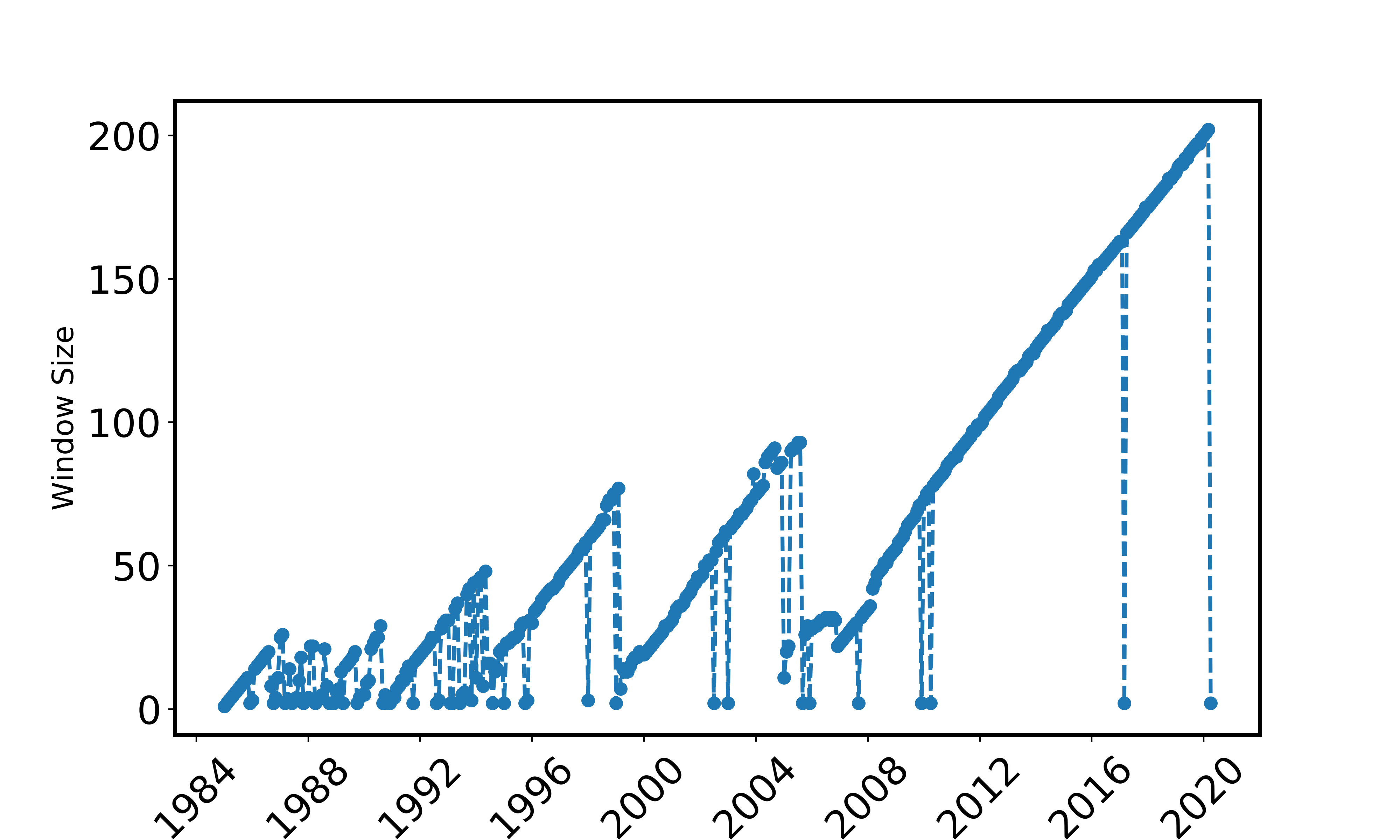}
}

\subfloat[]{
\includegraphics[width=.5\textwidth]{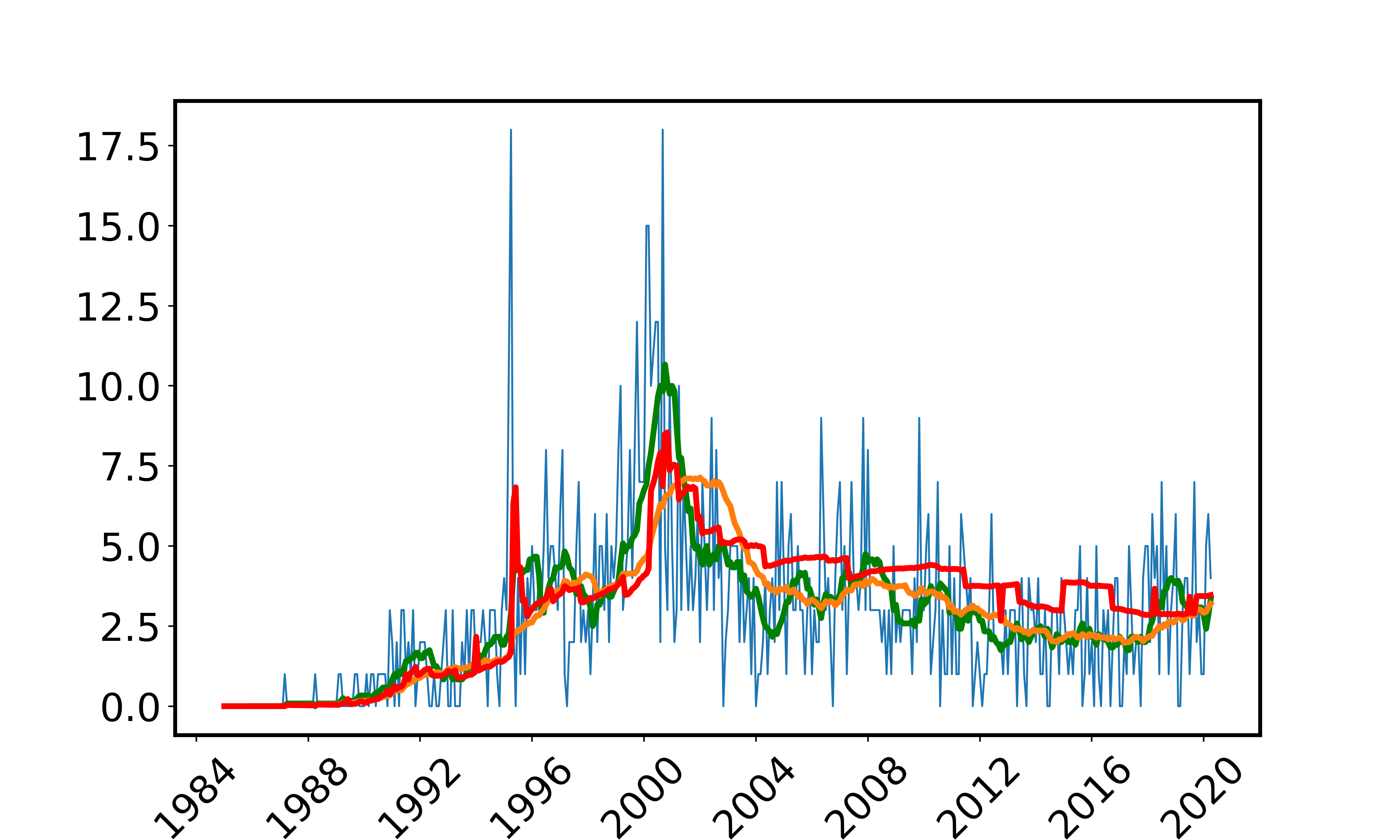}
\includegraphics[width=.5\textwidth]{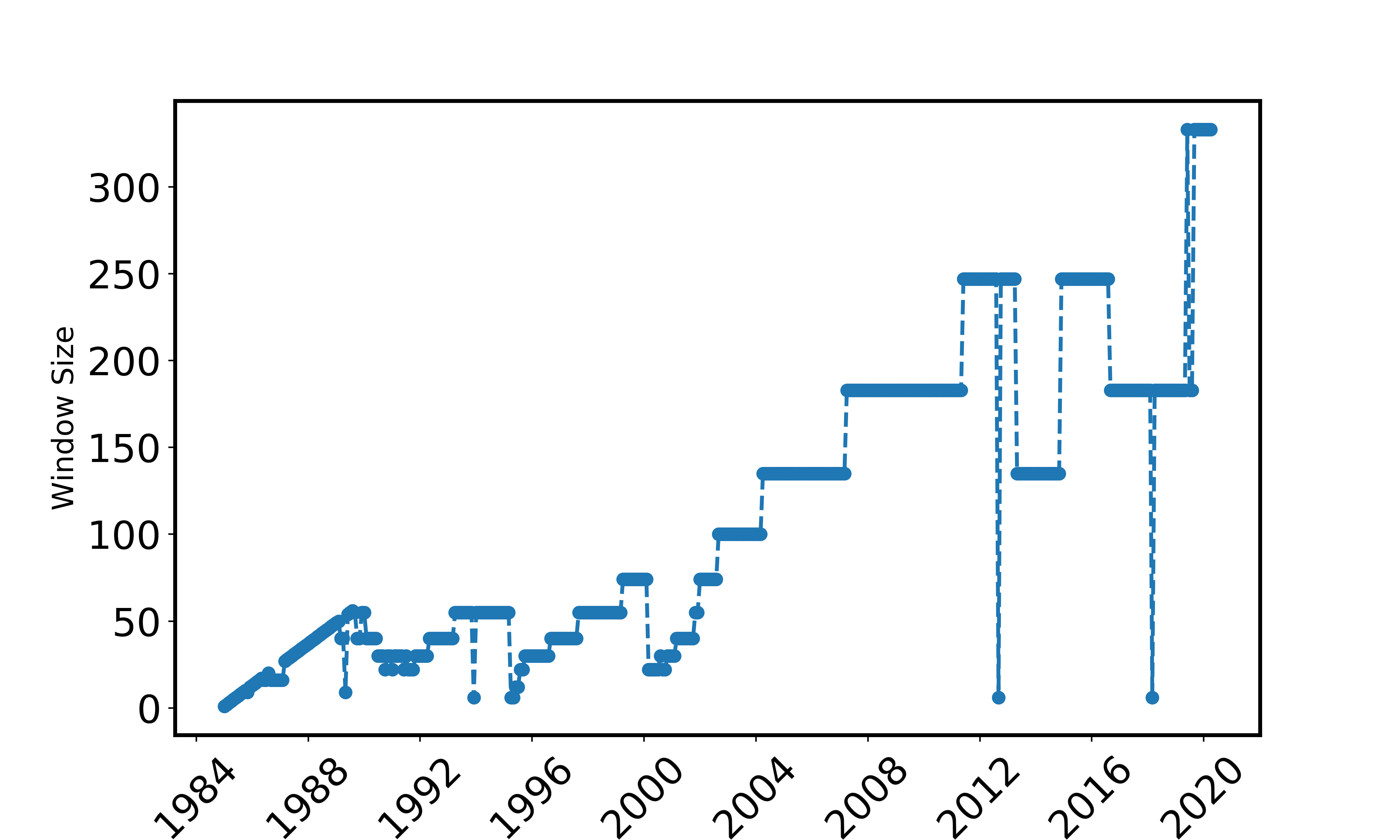}
}

\subfloat[]{
\includegraphics[width=.5\textwidth]{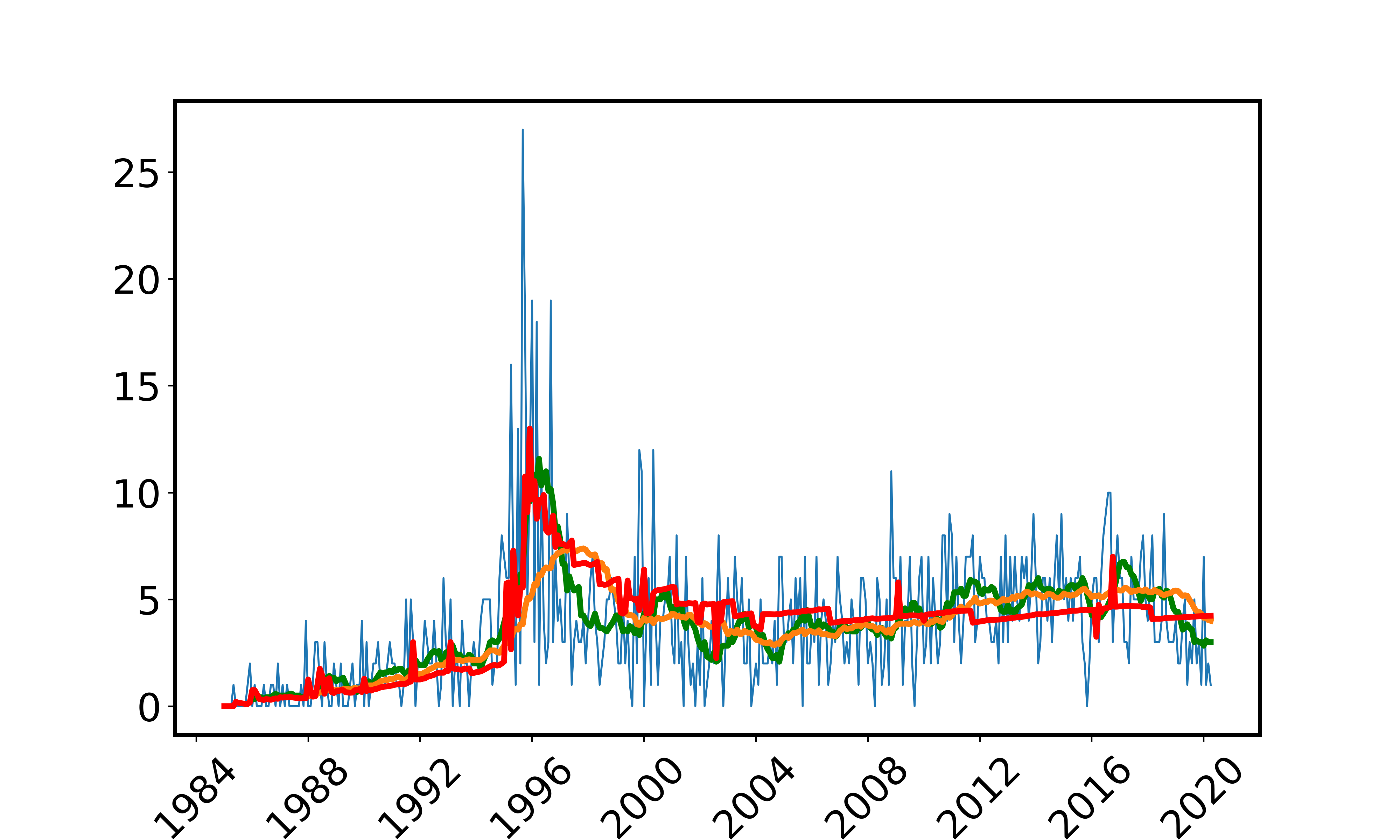}
\includegraphics[width=.5\textwidth]{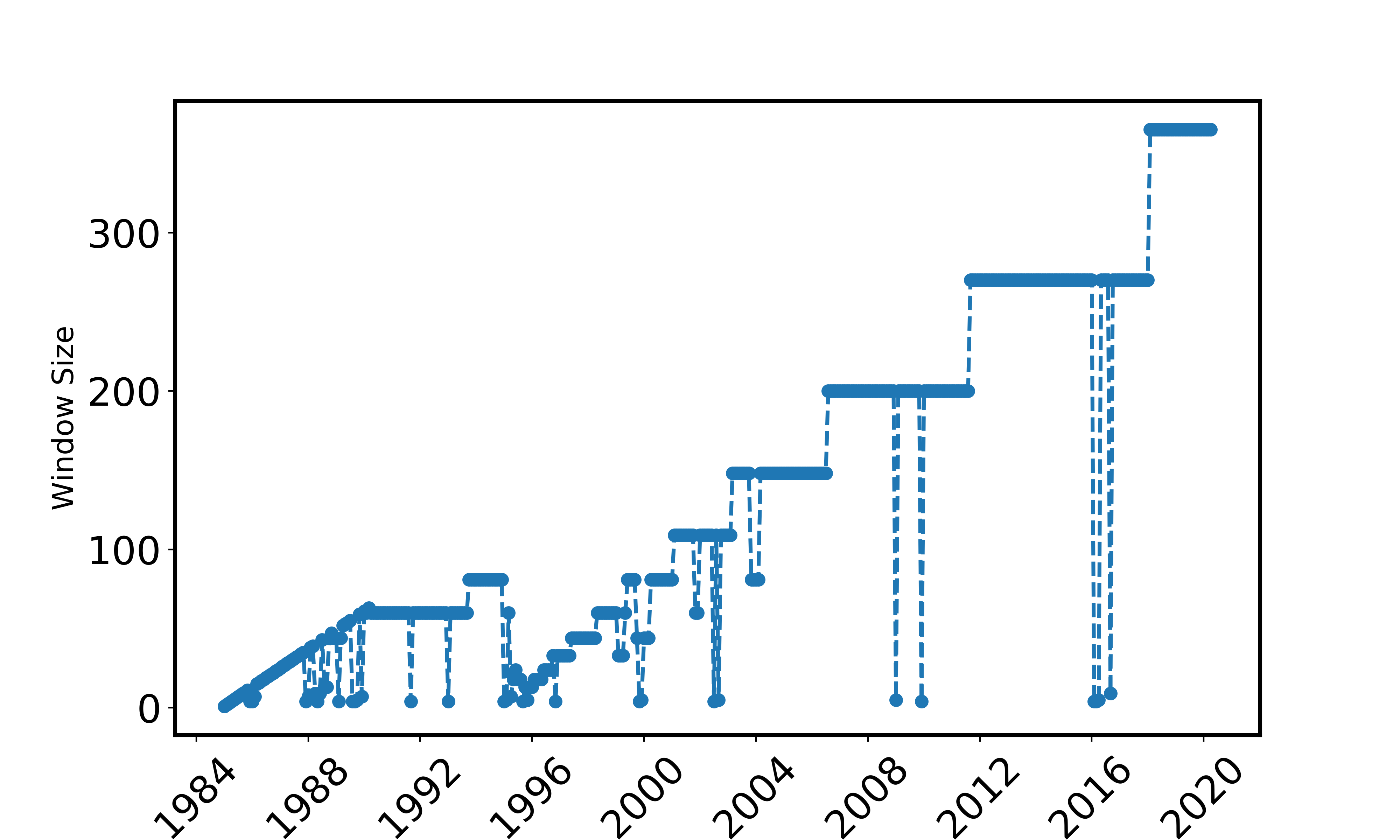}
}
\caption{{\color{sapphirecrayola} Time series of original data} and one step ahead prediction with estimates from {\color{red} LPA}, {\color{aoenglish}1 year fixed window} and {\color{orange} 3 year fixed window} (left) Homogeneous windows (right) for (1) Financials, (2) Telecommunication and (3) Energy.
\centering{ \protect \includegraphics[height=0.5cm]{images/qletlogo_tr.png} {\color{blue}\href{https://github.com/QuantLet/data_driven_controlling/tree/main/LPA_Empiricalstudy}{LPA\_Empiricalstudy}}}}
\label{fig:MA_results}
\end{figure}

Similar to the simulation study in previous section, we fix $n_0$ at 5 months, c at 1.35, compute K as in section (\ref{algohärdle}), and run 100 simulations to smoothen the estimated window sizes. We keep $n_0$ and c for the interval selection through equation (\ref{eq:interval_len}) constant as an aim of the proposed algorithm is to evaluate unseen data even if no knowledge about it is available. Some example cases are illustrated to verify the robustness of the algorithm with respect to different types of time series in figure (\ref{fig:MA_results}). The plot on the right hand side shows the homogeneous windows for each point in time while the corresponding LPA estimate, 12 month estimate and 36 month estimate as one period ahead forecast are shown in the left plot. 

\renewcommand{\thesubfigure}{\arabic{subfigure}}

The first plot in figure (\ref{fig:MA_results}) shows the German financial industry, where an upward trend in the number of mergers can be observed over time. The plot shows that LPA estimates closely mimic this trend, and is much more responsive to shocks as in the mid-90s. Moreover, contrary to the green 1-year moving average curve, which shows high variance in the MLE estimate, LPA recognizes the regions within time series where the average number of M\&A is approximately constant, such as from 2010 to 2020. The LPA estimates on the left suggest that the selected window size differs over time (sometimes up to 200, sometimes only a few observations) and using a fixed time window for generating forecasts is not recommended according to the technique. Comparing the window plot of the same industry with the simulation plots (a) and (b) from the previous section, we identify six regimes in the German financial industry (1984-1988, 1988-1990, 1990-1994, 1994-1999, 1999-2006 and 2006-2020). These regimes could be associated with external events, such as: a merger wave in 1984, globalization in late 1980s, the German bank merger wave in 1990s, German market liberalization in mid-90s, and the technological innovations in late 20th century.


The second part of figure (\ref{fig:MA_results}) shows the German telecommunication industry, where the number of M\&As per month seems stable and stationary, except for a short merger wave in the mid-90s due to the liberalization of the market and another short wave around 2001 (possibly related to the dotcom-bubble). LPA suggests a stable time series as the algorithm recommends to select large windows, and sometimes the whole time series. Only during the interval where a shock can be seen on the left, LPA's choice of homogeneous window on the right restricts the time window to a few years (for example between 1994 and 2001). We see that LPA can handle time-varying parameters while facing a trade-off between parameter variability and modelling bias. A similar conclusion can be drawn from the energy markets plots in (3) of figure (\ref{fig:MA_results}).

\begin{figure}[!htb]
\centering

\subfloat[]{
\includegraphics[width=.29\textwidth]{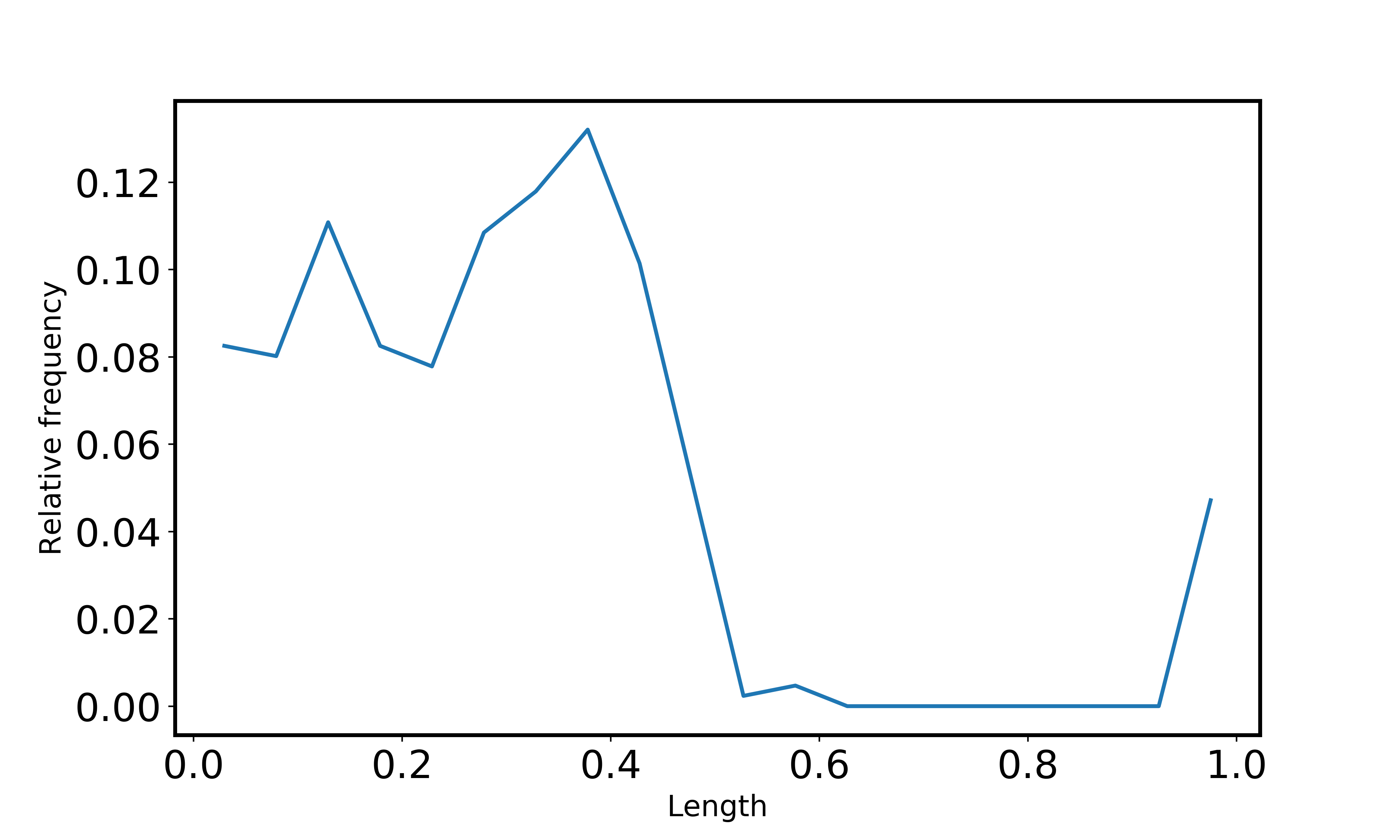}
}
\subfloat[]{
\includegraphics[width=.29\textwidth]{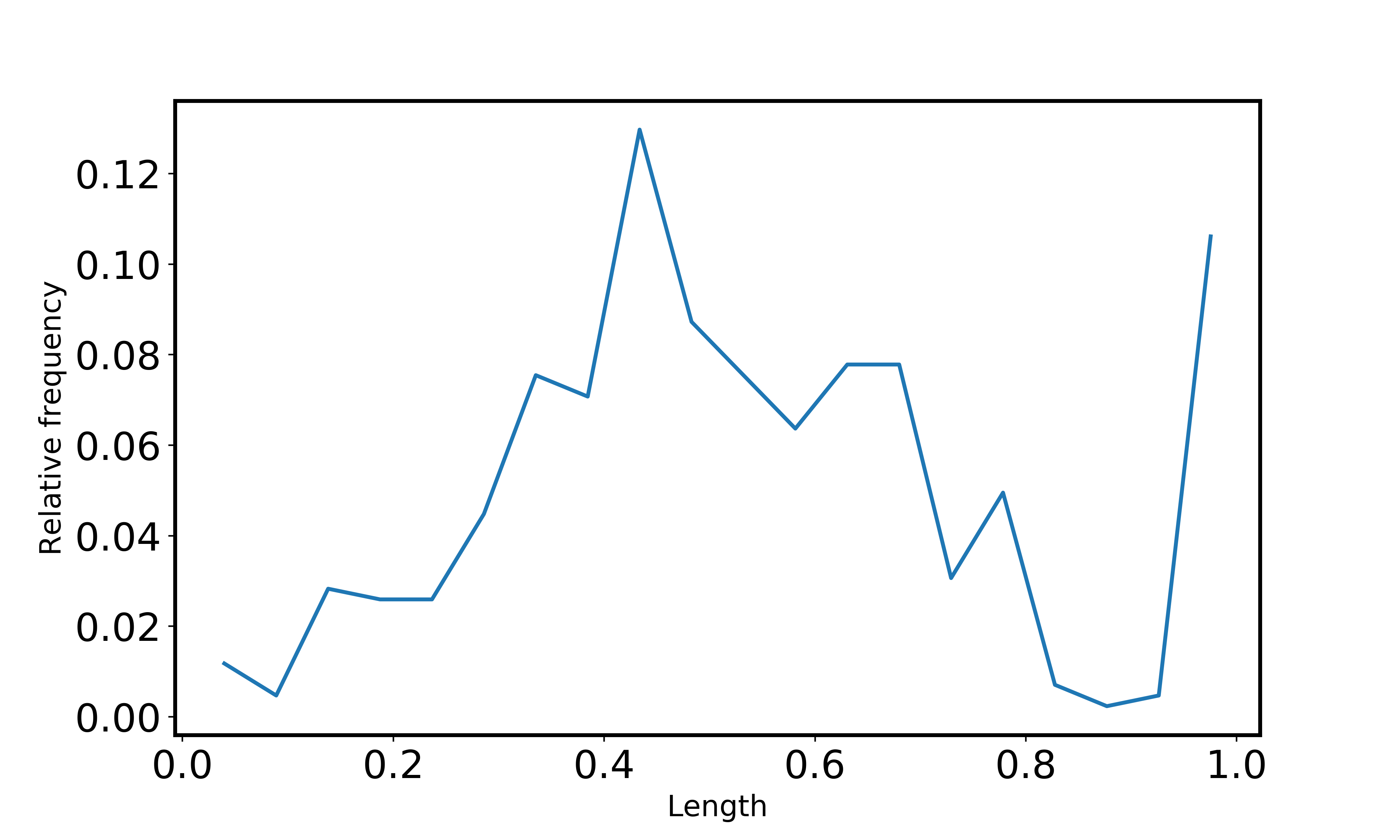}
}
\subfloat[]{
\includegraphics[width=.29\textwidth]{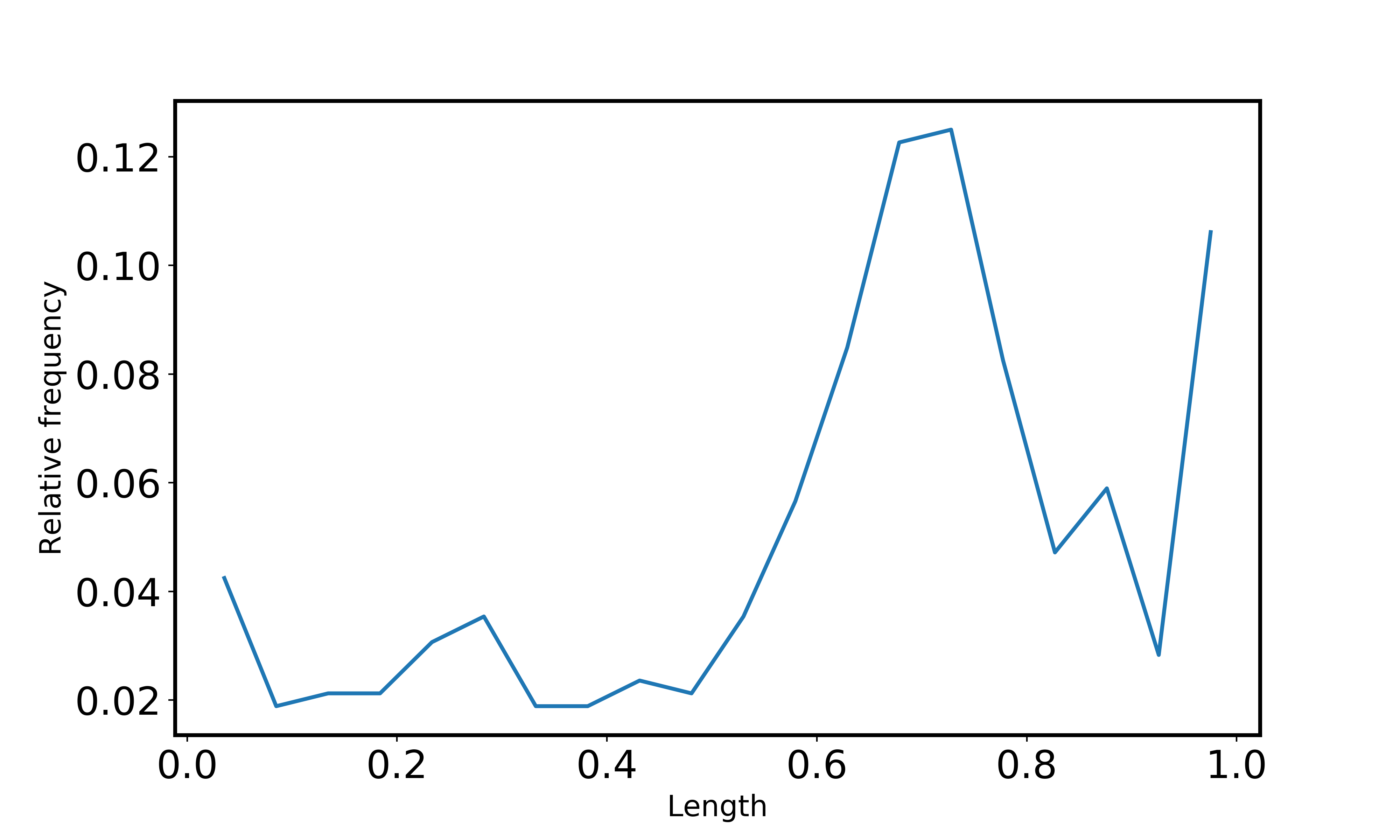}
}

\caption{Distribution of estimated interval length as a proportion of data points for (1) Financials, (2) Telecommunication and (3) Energy.\\
\centering{
\protect \includegraphics[height=0.5cm]{images/qletlogo_tr.png} {\color{blue}\href{https://github.com/QuantLet/data_driven_controlling/tree/main/LPA_Empiricalstudy}{LPA\_Empiricalstudy}}
}}
\label{fig:MA_mle_dist}
\end{figure}

Next, we forecast the number of M\&As for 02-2020 to 04-2020 using the LPA estimate and the fixed 1-year and 3-year window estimates of 01-2020. The estimates and mean squared errors (MSE) of estimation are summarized in the table (\ref{tab:mse}). Moreover, apart from the selection time-varying window lengths, we propose that LPA can also provide guidance on a priori selection of fixed windows. Looking at the distribution plot of window lengths (as a proportion of data points in the time series before the time of evaluation) in figure (\ref{fig:MA_mle_dist}) and selecting the window with highest frequency \((w)\) as the fixed time window for each time series, we also make a forecast and report the MSE in the same table. 

\begin{table}[!htb]
\centering
\begin{tabular}{llll}
                                 & Financials & Telecom. & Energy \\ \hline
LPA estimate                     & 52.99      & 3.43              & 4.23   \\
12 month MA estimate             & 55.58      & 2.42              & 3.08   \\
36 month MA estimate             & 55.50      & 2.94              & 4.17   \\
Most recurring window(w)        & 160        & 184               & 309    \\
w month MA estimate               & 55.25      & 2.90              & 4.62   \\
LPA MSE                          & 1008.24    & 3.11              & 8.61   \\
12 month MSE                     & 1159.29    & 7.34              & 3.29   \\
36 month MSE                     & 1154.25    & 4.89              & 8.25   \\
w month MSE                      & 1139.23    & 5.06              & 11.03\\  \hline
\end{tabular}
\caption{Forecast results for 02-2020 to 04-2020 based on the adaptively selected MLE, fixed windows and most recurring window proportionally \((w)\) in the distribution plots in figure (\ref{fig:MA_mle_dist})}
\label{tab:mse}
\end{table}

\renewcommand{\thesubfigure}{\arabic{subfigure}}

The table shows that LPA produces the smallest MSE in the financials and telecommunication industry. However, in the energy sector the 1 year fixed window outperforms LPA, perhaps due to its high fluctuations. Similarly, the homogeneous time window recommended by LPA for the financial industry was roughly 160 months. Choosing this time window resulted in a lower MSE than using 12 and 36 months fixed window estimates. For the energy sector, however, LPA recommends the selection of a very long window, for which the estimate deviates significantly from the true value, but still captures the stationary aspect of the time series. As a whole, the results show that wave-like patterns and level shifts are accurately detected, and that a plausible fit is achieved under the assumption that M\&A follow a Poisson distribution with time-varying parameter. 

The analysis shows that the proposed technique can be applied to real-world financial data and gives additional insights based merely on an automatic procedure. It can be used as a baseline for forecasting or simulation approaches. The MLE series could be forecasted using e.g. autoregressive models that generate different scenarios. Business experts could then make a judgement about the likeliness of these scenarios. Moreover, the availability of a time-varying parameter allows analysts to use more sophisticated approaches, e.g. forecasting densities of various time-series that are assumed to follow a Poisson distribution, such as the examples mentioned in the introduction (call centers, sales, supply chains). Businesses are often not interested in exact values, but in having an overview over the range of expected figures, and forecasts of low quality can lead to costly miscalculations. The proposed procedure could help increase forecast quality and availability and thus generate business value if employed.

\section{Conclusion}

A new algorithm for automatically finding homogeneous time intervals in a non stationary context is proposed. Trends, cyclical components, and even black swan events can be analysed with this technique. The algorithm is based on a fruitful combination of a local parametric model (here a Poisson process) and MBS. We conduct a simulation study that indicates that the procedure is indeed robust, even if the input data is not Poisson distributed. These results are then used to evaluate a data set of M\&A industry-based in Germany. The results show that in this non-parametric model-free context we obtain interpretable structural breaks and changes in trend. In conclusion, we provide an easy-to use solution for analyzing large data collections automatically that addresses the most common problems in time series modeling and can be adapted to diverse applications, including call center arrivals, sales forecasting or supply chain decisions.
\label{conclusion}

\section{Future Work}
\label{futurework}
This work is subject to several limitations that are caused either by computational cost or additional complexity that exceeds the scope of this work. As the proposed algorithm is computationally expensive, we evaluated only three German industries and admit that this choice is, to an extend, arbitrarily taken. However, data for all industries in the US and Germany is available and could be used. The simulations showed the the proposed method had difficulties with detecting small changes. We also saw that the method is not in all cases better than choosing a simple approach using moving averages. An evaluation based on all 20 datasets could give further insights on the robustness of the method. Furthermore, the evaluation using a pseudo-out-of-sample point forecast is fairly limited and we justify the superiority of our approach based on the simple measure of a mean-squared-error. However, as \cite{diebold_comparing_2015} outlines, this approach provides no insurance against over-fitting, are costly as they discard data and it is questionable whether they provide any benefit. Another questionable point is whether the occurrence of M\&A per month strictly follows a time-varying Poisson process, as assumed in this paper.

Answering this question using a sophisticated evaluation approach through generating (multi-period) density forecasts would give more insights and could be an interesting area for future research. Future research could also evaluate the fit to a Poisson distribution with time-varying MLE, but with respect to different values of $n_0$ and c. Alternatively, other methods for increasing the tested homogeneous interval could be compared to a geometric approach, e.g. random increases that follow a Poisson process or arithmetic increases. Furthermore, the simulation study indicated that the procedure could be generalized to work with any data that is assumed to follow a distribution of the exponential family. Finally, a possible extension could include a more general version of the current test statistic that looks into both directions.

\FloatBarrier



%
%

\bibliography{references.bib}

\begin{thebibliography}{41}
\providecommand{\natexlab}[1]{#1}
\providecommand{\url}[1]{\texttt{#1}}
\expandafter\ifx\csname urlstyle\endcsname\relax
  \providecommand{\doi}[1]{doi: #1}\else
  \providecommand{\doi}{doi: \begingroup \urlstyle{rm}\Url}\fi

\bibitem[Ahern and Harford(2014)]{ahern_importance_2014}
K.~R. Ahern and J.~Harford.
\newblock The {Importance} of {Industry} {Links} in {Merger} {Waves}.
\newblock \emph{The Journal of Finance}, 69\penalty0 (2):\penalty0 527--576,
  2014.
\newblock ISSN 1540-6261.
\newblock \doi{10.1111/jofi.12122}.
\newblock \_eprint: https://onlinelibrary.wiley.com/doi/pdf/10.1111/jofi.12122.

\bibitem[Akkus et~al.(2015)Akkus, Cookson, and Horta{\c
  c}su]{akkus_determinants_2015}
O.~Akkus, J.~A. Cookson, and A.~Horta{\c c}su.
\newblock The {Determinants} of {Bank} {Mergers}: {A} {Revealed} {Preference}
  {Analysis}.
\newblock \emph{Management Science}, 62\penalty0 (8):\penalty0 2241--2258, Nov.
  2015.
\newblock ISSN 0025-1909.
\newblock \doi{10.1287/mnsc.2015.2245}.
\newblock Publisher: INFORMS.

\bibitem[Aminikhanghahi and Cook(2017)]{aminikhanghahi_survey_2017}
S.~Aminikhanghahi and D.~J. Cook.
\newblock A {Survey} of {Methods} for {Time} {Series} {Change} {Point}
  {Detection}.
\newblock \emph{Knowledge and information systems}, 51\penalty0 (2):\penalty0
  339--367, May 2017.
\newblock ISSN 0219-1377.
\newblock \doi{10.1007/s10115-016-0987-z}.

\bibitem[Andrews and Ploberger(1994)]{andrews_optimal_1994}
D.~W.~K. Andrews and W.~Ploberger.
\newblock Optimal {Tests} when a {Nuisance} {Parameter} is {Present} {Only}
  {Under} the {Alternative}.
\newblock \emph{Econometrica}, 62\penalty0 (6):\penalty0 1383--1414, 1994.
\newblock ISSN 0012-9682.
\newblock \doi{10.2307/2951753}.
\newblock Publisher: [Wiley, Econometric Society].

\bibitem[Arlot et~al.(2010)Arlot, Blanchard, and
  Roquain]{arlot_nonasymptotic_2010}
S.~Arlot, G.~Blanchard, and E.~Roquain.
\newblock Some nonasymptotic results on resampling in high dimension, {I}:
  {Confidence} regions.
\newblock \emph{Annals of Statistics}, 38\penalty0 (1):\penalty0 51--82, Feb.
  2010.
\newblock ISSN 0090-5364, 2168-8966.
\newblock \doi{10.1214/08-AOS667}.
\newblock Publisher: Institute of Mathematical Statistics.

\bibitem[Beran(1986)]{beran_discussion_1986}
R.~Beran.
\newblock Discussion: {Jackknife}, {Bootstrap} and {Other} {Resampling}
  {Methods} in {Regression} {Analysis}.
\newblock \emph{Annals of Statistics}, 14\penalty0 (4):\penalty0 1295--1298,
  Dec. 1986.
\newblock ISSN 0090-5364, 2168-8966.
\newblock \doi{10.1214/aos/1176350143}.
\newblock Publisher: Institute of Mathematical Statistics.

\bibitem[B{\"u}cher and Dette(2013)]{bucher_multiplier_2013}
A.~B{\"u}cher and H.~Dette.
\newblock Multiplier bootstrap of tail copulas with applications.
\newblock \emph{Bernoulli}, 19\penalty0 (5A):\penalty0 1655--1687, Nov. 2013.
\newblock ISSN 1350-7265.
\newblock \doi{10.3150/12-BEJ425}.
\newblock Publisher: Bernoulli Society for Mathematical Statistics and
  Probability.

\bibitem[Chatterjee and Bose(2005)]{chatterjee_generalized_2005}
S.~Chatterjee and A.~Bose.
\newblock Generalized bootstrap for estimating equations.
\newblock \emph{Annals of Statistics}, 33\penalty0 (1):\penalty0 414--436, Feb.
  2005.
\newblock ISSN 0090-5364, 2168-8966.
\newblock \doi{10.1214/009053604000000904}.
\newblock Publisher: Institute of Mathematical Statistics.

\bibitem[Chen and Gupta(1999)]{chen_change_1999}
J.~Chen and A.~K. Gupta.
\newblock Change point analysis of a {Gaussian} model.
\newblock \emph{Statistical Papers}, 40\penalty0 (3):\penalty0 323--333, Sept.
  1999.
\newblock ISSN 1613-9798.
\newblock \doi{10.1007/BF02929878}.

\bibitem[Chen and Gupta(2011)]{chen_parametric_2011}
J.~Chen and A.~K. Gupta.
\newblock \emph{Parametric {Statistical} {Change} {Point} {Analysis}: {With}
  {Applications} to {Genetics}, {Medicine}, and {Finance}}.
\newblock Springer Science \& Business Media, Nov. 2011.
\newblock ISBN 978-0-8176-4801-5.
\newblock Google-Books-ID: mwzCuRMUVLIC.

\bibitem[Chernozhukov et~al.(2013)Chernozhukov, Chetverikov, and
  Kato]{chernozhukov_gaussian_2013}
V.~Chernozhukov, D.~Chetverikov, and K.~Kato.
\newblock Gaussian approximations and multiplier bootstrap for maxima of sums
  of high-dimensional random vectors.
\newblock \emph{Annals of Statistics}, 41\penalty0 (6):\penalty0 2786--2819,
  Dec. 2013.
\newblock ISSN 0090-5364, 2168-8966.
\newblock \doi{10.1214/13-AOS1161}.
\newblock Publisher: Institute of Mathematical Statistics.

\bibitem[Diebold(2015)]{diebold_comparing_2015}
F.~X. Diebold.
\newblock Comparing {Predictive} {Accuracy}, {Twenty} {Years} {Later}: {A}
  {Personal} {Perspective} on the {Use} and {Abuse} of
  {Diebold}{\textendash}{Mariano} {Tests}.
\newblock \emph{Journal of Business \& Economic Statistics}, 33\penalty0
  (1):\penalty0 1--1, Jan. 2015.
\newblock ISSN 0735-0015.
\newblock \doi{10.1080/07350015.2014.983236}.
\newblock Publisher: Taylor \& Francis.

\bibitem[Diebold et~al.(1998)Diebold, Gunther, and
  Tay]{diebold_evaluating_1998}
F.~X. Diebold, T.~A. Gunther, and A.~S. Tay.
\newblock Evaluating {Density} {Forecasts} with {Applications} to {Financial}
  {Risk} {Management}.
\newblock \emph{International Economic Review}, 39\penalty0 (4):\penalty0
  863--883, 1998.
\newblock ISSN 0020-6598.
\newblock \doi{10.2307/2527342}.
\newblock Publisher: [Economics Department of the University of Pennsylvania,
  Wiley, Institute of Social and Economic Research, Osaka University].

\bibitem[Eckley et~al.(2011)Eckley, Fearnhead, and
  Killick]{eckley_analysis_2011}
I.~A. Eckley, P.~Fearnhead, and R.~Killick.
\newblock Analysis of changepoint models, Aug. 2011.
\newblock Library Catalog: www.cambridge.org Pages: 205-224 Publisher:
  Cambridge University Press.

\bibitem[Giraitis et~al.(2013)Giraitis, Kapetanios, and
  Price]{giraitis_adaptive_2013}
L.~Giraitis, G.~Kapetanios, and S.~Price.
\newblock Adaptive forecasting in the presence of recent and ongoing structural
  change.
\newblock \emph{Journal of Econometrics}, 177\penalty0 (2):\penalty0 153--170,
  Dec. 2013.
\newblock ISSN 0304-4076.
\newblock \doi{10.1016/j.jeconom.2013.04.003}.

\bibitem[Gonzalez-Rivera and Sun(2014)]{gonzalez-rivera_density_2014}
G.~Gonzalez-Rivera and Y.~Sun.
\newblock Density {Forecast} {Evaluation} in {Unstable} {Environments}.
\newblock Technical Report 201428, University of California at Riverside,
  Department of Economics, Aug. 2014.
\newblock Publication Title: Working Papers.

\bibitem[Haccou et~al.(1987)Haccou, Meelis, and van~de
  Geer]{haccou_likelihood_1987}
P.~Haccou, E.~Meelis, and S.~van~de Geer.
\newblock The likelihood ratio test for the change point problem for
  exponentially distributed random variables.
\newblock \emph{Stochastic Processes and their Applications}, 27:\penalty0
  121--139, Jan. 1987.
\newblock ISSN 0304-4149.
\newblock \doi{10.1016/0304-4149(87)90009-3}.

\bibitem[H{\"a}rdle and Mammen(1993)]{hardle_comparing_1993}
W.~K. H{\"a}rdle and E.~Mammen.
\newblock Comparing {Nonparametric} {Versus} {Parametric} {Regression} {Fits}.
\newblock \emph{Annals of Statistics}, 21\penalty0 (4):\penalty0 1926--1947,
  Dec. 1993.
\newblock ISSN 0090-5364, 2168-8966.
\newblock \doi{10.1214/aos/1176349403}.
\newblock Publisher: Institute of Mathematical Statistics.

\bibitem[H{\"a}rdle et~al.(2014)H{\"a}rdle, Mihoci, and
  Hian-Ann~Ting]{hardle_adaptive_2014}
W.~K. H{\"a}rdle, A.~Mihoci, and C.~Hian-Ann~Ting.
\newblock Adaptive {Order} {Flow} {Forecasting} with {Multiplicative} {Error}
  {Models}.
\newblock {SSRN} {Scholarly} {Paper} ID 2892620, Social Science Research
  Network, Rochester, NY, July 2014.

\bibitem[H{\"a}rdle et~al.(2015)H{\"a}rdle, Hautsch, and
  Mihoci]{hardle_local_2015}
W.~K. H{\"a}rdle, N.~Hautsch, and A.~Mihoci.
\newblock Local {Adaptive} {Multiplicative} {Error} {Models} for
  {High}-{Frequency} {Forecasts}.
\newblock \emph{Journal of Applied Econometrics}, 30\penalty0 (4):\penalty0
  529--550, 2015.
\newblock ISSN 1099-1255.
\newblock \doi{10.1002/jae.2376}.
\newblock \_eprint: https://onlinelibrary.wiley.com/doi/pdf/10.1002/jae.2376.

\bibitem[Harford(2005)]{harford_what_2005}
J.~Harford.
\newblock What drives merger waves?
\newblock \emph{Journal of Financial Economics}, 77\penalty0 (3):\penalty0
  529--560, Sept. 2005.
\newblock ISSN 0304-405X.
\newblock \doi{10.1016/j.jfineco.2004.05.004}.

\bibitem[Hinkley and Hinkley(1970)]{hinkley_inference_1970}
D.~V. Hinkley and E.~A. Hinkley.
\newblock Inference {About} the {Change}-{Point} in a {Sequence} of {Binomial}
  {Variables}.
\newblock \emph{Biometrika}, 57\penalty0 (3):\penalty0 477--488, 1970.
\newblock ISSN 0006-3444.
\newblock \doi{10.2307/2334766}.
\newblock Publisher: [Oxford University Press, Biometrika Trust].

\bibitem[Hsu(1979)]{hsu_detecting_1979}
D.~A. Hsu.
\newblock Detecting {Shifts} of {Parameter} in {Gamma} {Sequences} with
  {Applications} to {Stock} {Price} and {Air} {Traffic} {Flow} {Analysis}.
\newblock \emph{Journal of the American Statistical Association}, 74\penalty0
  (365):\penalty0 31--40, Mar. 1979.
\newblock ISSN 0162-1459.
\newblock \doi{10.1080/01621459.1979.10481604}.
\newblock Publisher: Taylor \& Francis \_eprint:
  https://www.tandfonline.com/doi/pdf/10.1080/01621459.1979.10481604.

\bibitem[Inoue et~al.(2017)Inoue, Jin, and Rossi]{inoue_rolling_2017}
A.~Inoue, L.~Jin, and B.~Rossi.
\newblock Rolling window selection for out-of-sample forecasting with
  time-varying parameters.
\newblock \emph{Journal of Econometrics}, 196\penalty0 (1):\penalty0 55--67,
  Jan. 2017.
\newblock ISSN 0304-4076.
\newblock \doi{10.1016/j.jeconom.2016.03.006}.

\bibitem[Klochkov et~al.(2019)Klochkov, H{\"a}rdle, and
  Xu]{klochkov_localizing_2019}
Y.~Klochkov, W.~K. H{\"a}rdle, and X.~Xu.
\newblock Localizing {Multivariate} {CAViaR}.
\newblock \emph{IRTG1792 Discussion paper}, page~48, 2019.

\bibitem[Kolsarici and Vakratsas(2015)]{kolsarici_correcting_2015}
C.~Kolsarici and D.~Vakratsas.
\newblock Correcting for {Misspecification} in {Parameter} {Dynamics} to
  {Improve} {Forecast} {Accuracy} with {Adaptively} {Estimated} {Models}.
\newblock \emph{Management Science}, 61\penalty0 (10):\penalty0 2495--2513,
  Jan. 2015.
\newblock ISSN 0025-1909.
\newblock \doi{10.1287/mnsc.2014.2027}.
\newblock Publisher: INFORMS.

\bibitem[Kutoyants and Spokoiny(1999)]{kutoyants_optimal_1999}
Y.~A. Kutoyants and V.~Spokoiny.
\newblock Optimal choice of observation window for {Poisson} observations.
\newblock \emph{Statistics \& Probability Letters}, 44\penalty0 (3):\penalty0
  291--298, Sept. 1999.
\newblock ISSN 0167-7152.
\newblock \doi{10.1016/S0167-7152(99)00020-6}.

\bibitem[Maksimovic et~al.(2013)Maksimovic, Phillips, and
  Yang]{maksimovic_private_2013}
V.~Maksimovic, G.~Phillips, and L.~Yang.
\newblock Private and {Public} {Merger} {Waves}.
\newblock \emph{The Journal of Finance}, 68\penalty0 (5):\penalty0 2177--2217,
  2013.
\newblock ISSN 1540-6261.
\newblock \doi{10.1111/jofi.12055}.
\newblock \_eprint: https://onlinelibrary.wiley.com/doi/pdf/10.1111/jofi.12055.

\bibitem[Mammen(1993)]{mammen_bootstrap_1993}
E.~Mammen.
\newblock Bootstrap and {Wild} {Bootstrap} for {High} {Dimensional} {Linear}
  {Models}.
\newblock \emph{Annals of Statistics}, 21\penalty0 (1):\penalty0 255--285, Mar.
  1993.
\newblock ISSN 0090-5364, 2168-8966.
\newblock \doi{10.1214/aos/1176349025}.
\newblock Publisher: Institute of Mathematical Statistics.

\bibitem[Martynova and Renneboog(2005)]{martynova_century_2005}
M.~Martynova and L.~Renneboog.
\newblock A {Century} of {Corporate} {Takeovers}: {What} {Have} {We} {Learned}
  and {Where} {Do} {We} {Stand}? (previous title: {The} {History} of {M}\&{A}
  {Activity} {Around} the {World}: {A} {Survey} of {Literature}).
\newblock {SSRN} {Scholarly} {Paper} ID 820984, Social Science Research
  Network, Rochester, NY, Oct. 2005.

\bibitem[Mercurio and Spokoiny(2004)]{mercurio_statistical_2004}
D.~Mercurio and V.~Spokoiny.
\newblock Statistical inference for time-inhomogeneous volatility models.
\newblock \emph{Annals of Statistics}, 32\penalty0 (2):\penalty0 577--602, Apr.
  2004.
\newblock ISSN 0090-5364, 2168-8966.
\newblock \doi{10.1214/009053604000000102}.
\newblock Publisher: Institute of Mathematical Statistics.

\bibitem[Oreshkin et~al.(2016)Oreshkin, R{\'e}egnard, and
  L{\textquoteright}Ecuyer]{oreshkin_rate-based_2016}
B.~N. Oreshkin, N.~R{\'e}egnard, and P.~L{\textquoteright}Ecuyer.
\newblock Rate-{Based} {Daily} {Arrival} {Process} {Models} with {Application}
  to {Call} {Centers}.
\newblock \emph{Operations Research}, 64\penalty0 (2):\penalty0 510--527, Mar.
  2016.
\newblock ISSN 0030-364X.
\newblock \doi{10.1287/opre.2016.1484}.
\newblock Publisher: INFORMS.

\bibitem[Pesaran et~al.(2013)Pesaran, Pick, and
  Pranovich]{pesaran_optimal_2013}
M.~H. Pesaran, A.~Pick, and M.~Pranovich.
\newblock Optimal forecasts in the presence of structural breaks.
\newblock \emph{Journal of Econometrics}, 177\penalty0 (2):\penalty0 134--152,
  Dec. 2013.
\newblock ISSN 0304-4076.
\newblock \doi{10.1016/j.jeconom.2013.04.002}.

\bibitem[Spokoiny(2009)]{spokoiny_multiscale_2009}
V.~Spokoiny.
\newblock Multiscale local change point detection with applications to
  value-at-risk.
\newblock \emph{The Annals of Statistics}, 37\penalty0 (3):\penalty0
  1405--1436, June 2009.
\newblock ISSN 0090-5364, 2168-8966.
\newblock \doi{10.1214/08-AOS612}.
\newblock Publisher: Institute of Mathematical Statistics.

\bibitem[Spokoiny and Zhilova(2015)]{spokoiny_bootstrap_2015}
V.~Spokoiny and M.~Zhilova.
\newblock Bootstrap confidence sets under model misspecification.
\newblock \emph{Annals of Statistics}, 43\penalty0 (6):\penalty0 2653--2675,
  Dec. 2015.
\newblock ISSN 0090-5364, 2168-8966.
\newblock \doi{10.1214/15-AOS1355}.
\newblock Publisher: Institute of Mathematical Statistics.

\bibitem[Spokoiny(1998)]{spokoiny_estimation_1998}
V.~G. Spokoiny.
\newblock Estimation of a function with discontinuities via local polynomial
  fit with an adaptive window choice.
\newblock \emph{The Annals of Statistics}, 26\penalty0 (4):\penalty0
  1356--1378, Aug. 1998.
\newblock ISSN 0090-5364, 2168-8966.
\newblock \doi{10.1214/aos/1024691246}.
\newblock Publisher: Institute of Mathematical Statistics.

\bibitem[Taylor(2007)]{taylor_comparison_2007}
J.~W. Taylor.
\newblock A {Comparison} of {Univariate} {Time} {Series} {Methods} for
  {Forecasting} {Intraday} {Arrivals} at a {Call} {Center}.
\newblock \emph{Management Science}, 54\penalty0 (2):\penalty0 253--265, Dec.
  2007.
\newblock ISSN 0025-1909.
\newblock \doi{10.1287/mnsc.1070.0786}.
\newblock Publisher: INFORMS.

\bibitem[Taylor(2011)]{taylor_density_2011}
J.~W. Taylor.
\newblock Density {Forecasting} of {Intraday} {Call} {Center} {Arrivals}
  {Using} {Models} {Based} on {Exponential} {Smoothing}.
\newblock \emph{Management Science}, 58\penalty0 (3):\penalty0 534--549, Oct.
  2011.
\newblock ISSN 0025-1909.
\newblock \doi{10.1287/mnsc.1110.1434}.
\newblock Publisher: INFORMS.

\bibitem[Very et~al.(2012)Very, Metais, Lo, and Hourquet]{very_can_2012}
P.~Very, E.~Metais, S.~Lo, and P.-G. Hourquet.
\newblock Can {We} {Predict} {M}\&{A} {Activity}?
\newblock In S.~Finkelstein and C.~L.~Cooper, editors, \emph{Advances in
  {Mergers} and {Acquisitions}}, volume~11 of \emph{Advances in {Mergers} \&
  {Acquisitions}}, pages 1--32. Emerald Group Publishing Limited, Jan. 2012.
\newblock ISBN 978-1-78190-460-2 978-1-78190-459-6.
\newblock \doi{10.1108/S1479-361X(2012)0000011004}.

\bibitem[Wu(1986)]{wu_jackknife_1986}
C.~F.~J. Wu.
\newblock Jackknife, {Bootstrap} and {Other} {Resampling} {Methods} in
  {Regression} {Analysis}.
\newblock \emph{Annals of Statistics}, 14\penalty0 (4):\penalty0 1261--1295,
  Dec. 1986.
\newblock ISSN 0090-5364, 2168-8966.
\newblock \doi{10.1214/aos/1176350142}.
\newblock Publisher: Institute of Mathematical Statistics.

\bibitem[Yelland et~al.(2010)Yelland, Kim, and
  Stratulate]{yelland_bayesian_2010}
P.~M. Yelland, S.~Kim, and R.~Stratulate.
\newblock A {Bayesian} {Model} for {Sales} {Forecasting} at {Sun}
  {Microsystems}.
\newblock \emph{Interfaces}, 40\penalty0 (2):\penalty0 118--129, 2010.
\newblock ISSN 0092-2102.
\newblock Publisher: INFORMS.

\end{thebibliography}
\end{document}